\documentclass[twocolumn,showpacs,aps,prd,altaffilletter]{revtex4-1}

\usepackage{graphicx}
\usepackage{dcolumn}
\usepackage{amsmath}
\usepackage{epsfig}
\usepackage{verbatim}
\usepackage{mathrsfs}

\input babarsym

\newcommand{\BABARPubYear}    {14}
\newcommand{\BABARPubNumber}  {009}

\newcommand{\SLACPubNumber} {16180}
\newcommand{\LANLNumber} {1412.6751}

\def\figurebox#1#2#3{%
    \def\arg{#3}%
    \ifx\arg\empty
    {\hfill\vbox{\hsize#2\hrule\hbox to #2{\vrule\hfill\vbox to #1{\hsize#2\vfill}\vrule}\hrule}\hfill}%
    \else
    {\hfill\epsfbox{#3}\hfill}%
    \fi}

\def\babar{\mbox{\slshape B\kern-0.1em{\smaller A}\kern-0.1em
    B\kern-0.1em{\smaller A\kern-0.2em R}}\xspace}

\def\DsOneRes {\ensuremath{D_{s1}(2536)}\xspace}
\def\DsJRes {\ensuremath{D^{*}_{s1}(2700)^+}\xspace}
\def\DsTwo {\ensuremath{D^*_{s2}(2573)^+}\xspace}
\def\DsJTwoEight {\ensuremath{D_{sJ}^*(2860)^+}\xspace}
\def\DsJThreeZero {\ensuremath{D_{sJ}(3040)^+}\xspace}
\def\PsiRes {\ensuremath{\psi(3770)}\xspace}
\def\XRes {\ensuremath{X(3872)}\xspace}

\newcommand{\BToDDK}{\ensuremath{B\to \Dbar^{(*)} D^{(*)} K}\xspace}

\def\to {\rightarrow}

\def\modei{\ensuremath{\Bz\to \Dm \Dz \Kp}\xspace}

\def\modexi{\ensuremath{\Bu\to \Dzb \Dz \Kp}\xspace}

\def\mes    {\mbox{$m_{\rm ES}$}\xspace}
\def\de    {\mbox{$\Delta E$}\xspace}

\def\ndof {\ensuremath{n_{\mathrm{dof}}}}

\newcommand\CellTop{\rule{0pt}{2.35ex}}
\newcommand\CellTopTwo{\rule{0pt}{2.8ex}}
\newcommand\CellTopThree{\rule{0pt}{3.2ex}}

\begin{document}

\preprint{\babar-PUB-\BABARPubYear/\BABARPubNumber}
\preprint{SLAC-PUB-\SLACPubNumber}

\begin{flushleft}
\babar-PUB-\BABARPubYear/\BABARPubNumber \\
SLAC-PUB-\SLACPubNumber \\
arXiv:\LANLNumber\ [hep-ex]\\
\end{flushleft}

\title{
{\large \bf Dalitz plot analyses of \modei and \modexi decays} }

%
\author{J.~P.~Lees}
\author{V.~Poireau}
\author{V.~Tisserand}
\affiliation{Laboratoire d'Annecy-le-Vieux de Physique des Particules (LAPP), Universit\'e de Savoie, CNRS/IN2P3,  F-74941 Annecy-Le-Vieux, France}
\author{E.~Grauges}
\affiliation{Universitat de Barcelona, Facultat de Fisica, Departament ECM, E-08028 Barcelona, Spain }
\author{A.~Palano$^{ab}$ }
\affiliation{INFN Sezione di Bari$^{a}$; Dipartimento di Fisica, Universit\`a di Bari$^{b}$, I-70126 Bari, Italy }
\author{G.~Eigen}
\author{B.~Stugu}
\affiliation{University of Bergen, Institute of Physics, N-5007 Bergen, Norway }
\author{D.~N.~Brown}
\author{L.~T.~Kerth}
\author{Yu.~G.~Kolomensky}
\author{M.~J.~Lee}
\author{G.~Lynch}
\affiliation{Lawrence Berkeley National Laboratory and University of California, Berkeley, California 94720, USA }
% ...
\collaboration{The \babar\ Collaboration}
\noaffiliation

\begin{abstract}
\noindent
We present Dalitz plot analyses for
the decays of $B$ mesons to $D^- D^0 K^+$ and $\Dzb D^0 K^+$. We report the observation of the \DsJRes resonance in these two channels and obtain measurements of the mass $M(\DsJRes) = 2699 {}^{+14}_{-7} \mevcc$ and of the width $\Gamma(\DsJRes) = 127 {}^{+24}_{-19} \mev$, including statistical and systematic uncertainties. In addition, we observe an enhancement in the $D^0K^+$ invariant mass around 2350--2500 \mevcc in both decays \modei and \modexi, which we are not able to interpret. The results are based on $429~\mathrm{fb}^{-1}$ of data containing $471\times10^6\, B\Bbar$ pairs collected at the $\FourS$ resonance with the \babar\ detector at the SLAC National Accelerator Laboratory.
\end{abstract}


\maketitle


\section{Introduction}
\label{sec:intro}

\noindent
In the $\BToDDK$ final states~\cite{cc}, where $D$ is a \Dz or a \Dp, $D^*$ is a \Dstarz or \Dstarp, and $K$ is a \Kp or a $K^0$, we have the possibility to search for $\Dbar^{(*)} D^{(*)}$ resonances (charmonium or charmoniumlike resonances) as well as $D^{(*)} K$ resonances ($c \bar{s}$ resonances). These final states have been useful in the past to determine properties of the \DsOneRes and \PsiRes mesons and of the \XRes state at \babar~\cite{ref:DDKRes} and Belle~\cite{ref:Ds2536Belle,ref:DsJBelle} as well as the \DsJRes meson at Belle~\cite{ref:DsJBelle}. These analyses, based on the studies of invariant mass distributions, were able to extract properties such as the mass, width, and spin of the resonances.

The analysis presented here gives useful information about $c \bar{s}$ mesons present in these decays. Before 2003 only four $c \bar{s}$ mesons were known and their properties were consistent with the predictions of potential models~\cite{ref:godfrey}. Since then the $D^*_{s0}(2317)$ and $D_{s1}(2460)$ states have been discovered by \babar and CLEO~\cite{ref:Ds0_1} with widths and masses in disagreement with the expectations. The \DsJRes meson was discovered by \babar decaying to $DK$ in inclusive $e^+e^-$ interactions~\cite{ref:DsJAntimo}, and confirmed by Belle in the final state \modexi~\cite{ref:DsJBelle}. The LHCb experiment studied the \DsJRes meson in the $D^+\KS$ and $D^0K^+$ final states~\cite{ref:DsJLHCb} and obtained a more precise determination of its properties. The \DsJRes meson was also observed in inclusive $e^+e^-$ interactions decaying to $D^*K$~\cite{ref:DsJAntimo}. An additional $c \bar{s}$ state, the \DsJTwoEight, was discovered by \babar decaying to $DK$ and $D^*K$~\cite{ref:DsJAntimo}, and confirmed by LHCb~\cite{ref:DsJLHCb}. Recently, LHCb claimed that the structure at $2.86~\gevcc$ was an admixture of spin-1 and spin-3 resonances~\cite{ref:DsJ2860LHCb}. Finally, the \DsJThreeZero meson was observed in the $D^*K$ final state by \babar~\cite{ref:DsJAntimo}.

In this study, we perform Dalitz plot analyses for the channels \modei and \modexi which contain only pseudoscalar particles in the three-body decay. These Dalitz plot analyses allow the interferences between the different amplitudes which are present in the final states to be taken into account correctly. This is the first time that such Dalitz plot analyses have been performed for these decays.

\section{The \babar\ detector and data sample}

\noindent
The data were recorded by the \babar\ detector at the PEP-II asymmetric-energy $e^+e^-$ storage ring operating at the SLAC National Accelerator Laboratory. We analyze the complete \babar\ data sample collected at the $\FourS$ resonance corresponding to an integrated luminosity of 429~\invfb~\cite{ref:lumi}, giving $N_{\BB} = (470.9 \pm 0.1 \pm 2.8) \times 10^6$ \BB pairs produced, where the first uncertainty is statistical and the second systematic.

The \babar\ detector is described in detail elsewhere~\cite{ref:nim}. Charged particles are detected and their momenta measured with a five-layer silicon vertex tracker and a 40-layer drift chamber in a 1.5 T axial magnetic field. Charged particle identification is based on the measurements of the energy loss in the tracking devices and of the Cherenkov radiation in the ring-imaging detector. The energies and locations of showers associated with photons are measured in
the electromagnetic calorimeter. Muons are identified by the instrumented magnetic-flux return, which is located outside the magnet.

We employ a Monte Carlo simulation to study the relevant backgrounds and estimate the selection efficiencies.
We use \textsc{EVTGEN}~\cite{ref:evtgen} to model the kinematics of $B$ mesons and \textsc{JETSET}~\cite{ref:jetset} to model continuum processes, $\epem\to\qqbar$ ($q=u,d,s,c$). The \babar\ detector and its response to particle interactions are modeled using the \textsc{GEANT4}~\cite{ref:geant4} simulation package.

\section{Event selection and signal yields}
\label{sec:selection}

\noindent
The selection and reconstruction of \modei and \modexi, along with 20 other \BToDDK modes, is described in Ref.~\cite{ref:DDKBF}. We briefly summarize the selection criteria in this section. We reconstruct $D$ mesons in the modes $\Dz\to \Km\pip$, $\Km\pip\piz$, $\Km\pip\pim\pip$, and $\Dp\to \Km\pip\pip$. The $B$ candidates are reconstructed by combining a $\Dbar$ (representing either a $D^-$ or a $\Dzb$), a $D^0$, and a $K^+$ candidate. For the mode \modexi, at least one of the \Dz mesons is required to decay to $\Km \pip$. A mass-constrained kinematic fit is
applied to the intermediate particles (\Dz, \Dp, \piz) to improve their momentum resolution and the resolution of the invariant masses of the studied resonances.

Two kinematic variables are used to isolate the $B$-meson signal.
The first variable is the beam-energy-substituted mass defined as
\begin{equation}
\mes = \sqrt{{\left( {\frac {{s / 2}
+\vec{p}_0.\vec{p}_B} {E_0}}  \right)^2}-\vert \vec{p}_B \vert
^2},
\end{equation}
where $\sqrt{s}$ is the $e^+e^-$ center-of-mass energy. For the momenta in the laboratory frame,
$\vec{p}_0$, $\vec{p}_B$, and the energy, $E_0$, the subscripts $0$ and
$B$ refer to the $e^+e^-$ system and the reconstructed $B$ meson,
respectively. The other variable is $\de$, the difference between the reconstructed
energy of the $B$ candidate and the beam energy in the $e^+e^-$
center-of-mass frame. Signal events have \mes\ compatible with the known $B$-meson
mass~\cite{ref:pdg} and \de\ compatible with 0~\mev, within their respective resolutions.

For the modes \modei and \modexi, we obtain 1.1 and 1.3 $B$ candidates
per event on average, respectively. If more than one candidate is selected in an event, we retain only the one with the smallest value of $|\de|$. According to Monte Carlo studies, this criterion finds the correct candidate, when present in the candidate list, in more than 95\% of the cases for the two final states.
We keep only events with $|\de|<10-14 \mev$ depending on the $D$ final state~\cite{ref:DDKBF}.

\begin{figure}[htb]
\begin{center}
\epsfig{file=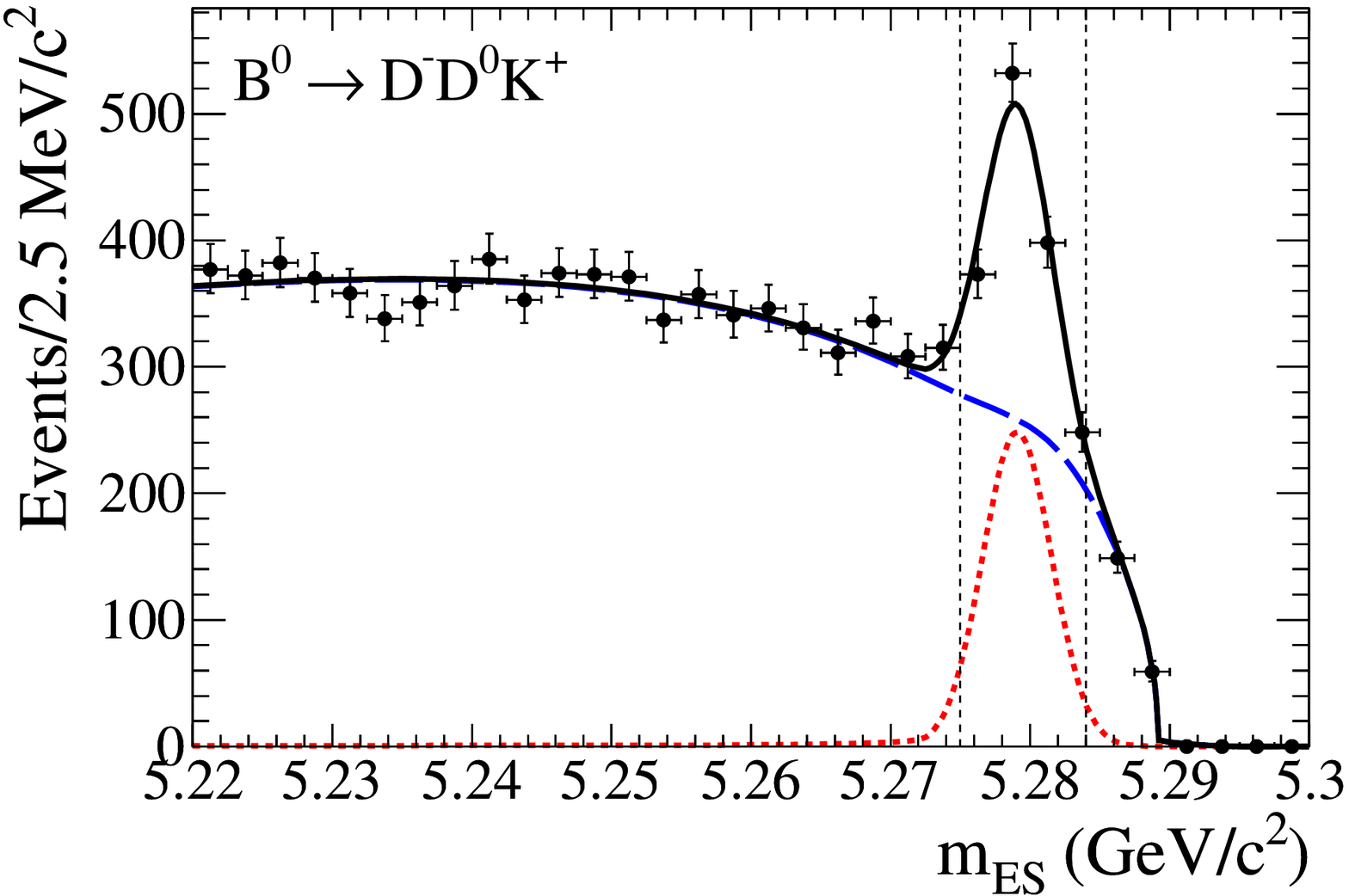,width=7.5cm}
\epsfig{file=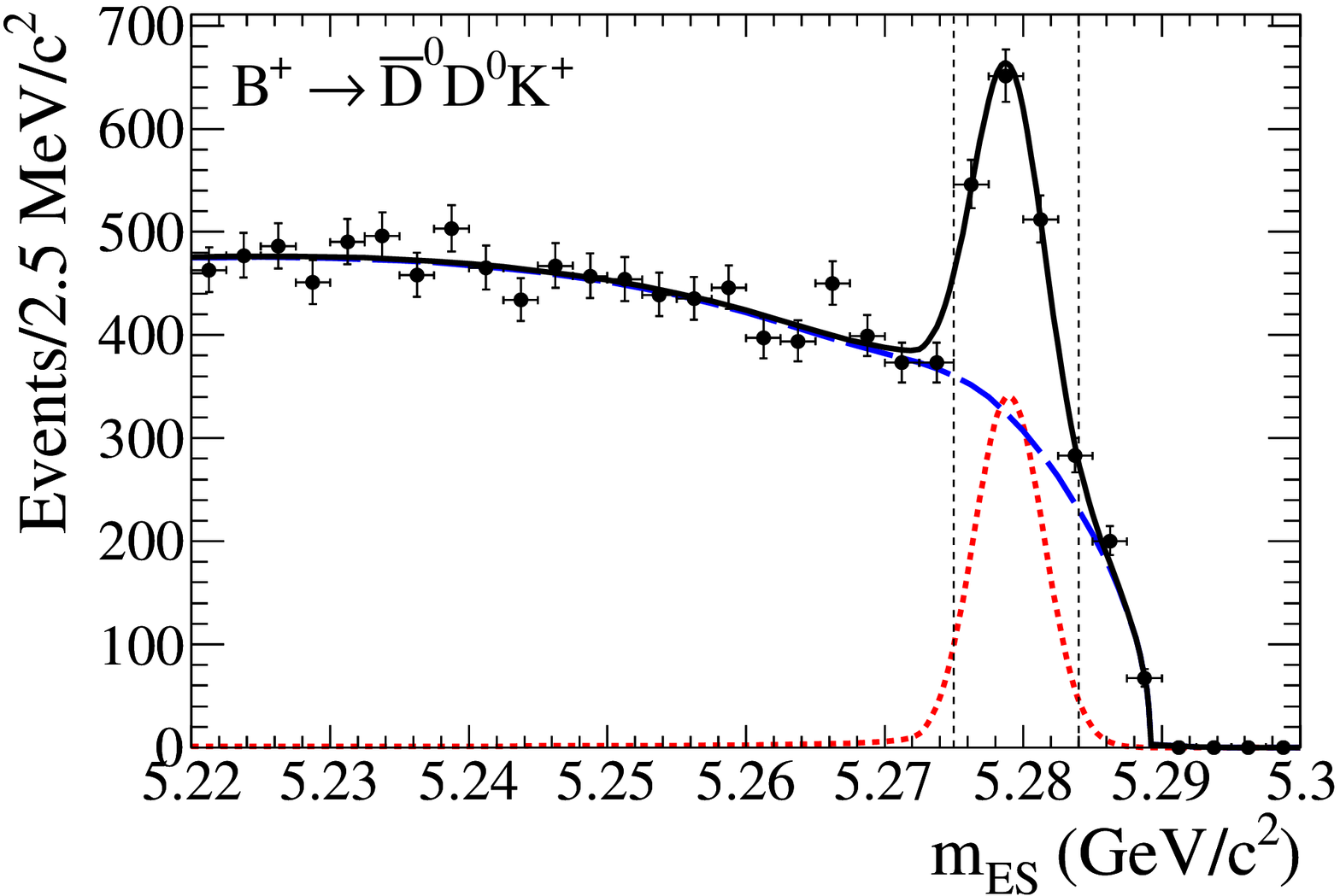,width=7.5cm}
 \caption{Fits of the \mes\ data distributions~\cite{ref:DDKBF} for the modes \modei (top) and \modexi (bottom).
 Points with statistical errors are data events, the red dashed line represents the signal PDF, the blue dashed line represents the background PDF, and the black solid line shows the total PDF. The vertical lines indicate the signal region used in the rest of the analysis.}
  \label{fig:mes}
\end{center}
\end{figure}

In order to obtain the signal and background yields, fits are performed on the \mes distributions, as described in detail in Ref.~\cite{ref:DDKBF}. The probability density function (PDF) of the signal is determined from Monte Carlo samples. A Crystal Ball function~\cite{ref:cb} (Gaussian PDF modified to include a power-law tail on the low side of the peak) is used to describe the signal. The background contribution is the sum of cross-feed events and combinatorial background.
The cross-feed background to a mode consists of all incorrectly reconstructed \BToDDK events. The ratio of
cross-feed events to signal events is 11\% for \modei and 17\% for \modexi.
The cross-feed events are described by the sum of an ARGUS function~\cite{ref:argus} and a Gaussian function, where the latter allows us to take into account the cross-feed events peaking in the signal region. For both modes, the Gaussian part represents a negligible contribution to the total cross feed. Since the fit for the yield for one channel uses as input the branching fractions from other channels, an iterative procedure using the 22 \BToDDK modes is employed~\cite{ref:DDKBF}.
The combinatorial background events are described by the sum of an ARGUS function and a Gaussian function, reflecting the fact that a small fraction of events peaks in the signal region (coming for example from $\Db D^0 K^+$ events where one of the $D$ mesons is not decaying to a studied mode).
The fits performed on the \mes distributions using the sum of the PDFs for the signal and for the background are shown in Fig.~\ref{fig:mes}. We obtain $635 \pm 47$ and $901 \pm 54$ signal events for \modei and \modexi, respectively~\cite{ref:DDKBF}.

We impose an additional condition to enhance the purity for the Dalitz plot analysis.
We require $5.275 < \mes < 5.284 \gevcc$, obtaining a total number of 1470 events with a signal purity of $(38.6 \pm 2.8 \pm 2.1)\%$ for \modei and obtaining a total of 1894 events with a signal purity of $(41.6 \pm 2.5 \pm 3.1)\%$ for \modexi, where the first uncertainties are statistical and the second systematic.

\section{Dalitz plot analyses}

\subsection{Method}

\noindent
We use an isobar model formalism to perform the Dalitz plot analysis~\cite{ref:cleoDalitz}.
The decays are described by a sum of amplitudes representing nonresonant and resonant contributions:
\begin{equation}
\mathcal{M} = \sum_i c_i A_i,
\end{equation}
where the $c_i \equiv \rho_i e^{i \phi_i}$ are complex coefficients of modulus $\rho_i \equiv |c_i|$ and phase $\phi_i$.
The quantities $A_i$ are complex amplitudes and can be written as:
\begin{equation}
\label{eq:Ai}
A_i = D_i \times T_i(\Omega),
\end{equation}
where $D_i$ represents the dynamical function describing the $i$th intermediate resonance,
and $T_i(\Omega)$ represents the angular distribution of the final state particles as a function of the solid angle $\Omega$. For nonresonant events, we have $A_i=1$. The quantity $D_i$ is parametrized by a Breit-Wigner function, whose expression for a resonance $r \to AB$ is given by
\begin{equation}
\label{eq:bw}
D = \frac{F_B F_r}{M_r^2-M^2_{AB}-i\Gamma_{AB}M_r},
\end{equation}
where $F_B$ and $F_r$ are the Blatt-Weisskopf damping factors for the $B$ meson and for the resonance, $M_r$ is the mass of the resonance, $M_{AB}$ the invariant mass of the system $AB$, and $\Gamma_{AB}$ its mass-dependent width. The expression for the mass-dependent width is
\begin{equation}
\Gamma_{AB} = \Gamma_r \left( \frac{p_{AB}}{p_r} \right)^{2J+1} \left( \frac{M_r}{M_{AB}} \right) F^2_r,
\end{equation}
where $\Gamma_r$ and $J$ are the width and the spin of the resonance. The quantity $p_{AB}$ is the momentum of either daughter in the $AB$ rest frame, and $p_r$ is the value of $p_{AB}$ when $M_{AB}=M_r$.
The resonances we study here have a large enough natural width and we can ignore the effect of the detector resolutions.

The exact expressions of the Blatt-Weisskopf factors~\cite{blatt} are given in Ref.~\cite{ref:antimo} and depend on the momenta of the particles involved, on the spin of the resonance, and on the radius of the Blatt-Weisskopf barrier. For the $B$ meson and for the intermediate resonance, we use a radius of $1.5 \gev^{-1}$.
The angular term $T_i(\Omega)$ is also given in Ref.~\cite{ref:antimo} and depends on the masses of the particles involved in the reaction as well as on the spin of the intermediate resonance.

We extract the complex amplitudes present in the data (from their modulus $\rho_i$ and phase $\phi_i$), and the mass
and width of the \DsJRes resonance. To obtain these parameters, we perform an unbinned maximum likelihood fit where the likelihood function for the event $n$ can be written as
\begin{eqnarray}
\label{eq:likelihood}
\mathcal{L}_n &=& p \times \varepsilon_i(m_1^2, m_2^2) \times \frac{|\mathcal{M}|^2}{\int |\mathcal{M}|^2 \varepsilon(m_1^2, m_2^2) dm_1^2 dm_2^2} \nonumber \\
&& + (1-p) \times \frac{B(m_1^2,m_2^2)}{\int B(m_1^2,m_2^2) dm_1^2 dm_2^2},
\end{eqnarray}
where $|\mathcal{M}|^2 = \sum_{i,j} c_i c_j^* A_i A_j^*$ is calculated for the event $n$. In this expression, $m_1$ and $m_2$ represent the invariant mass of $\Dbar D^0$ and $D^0K^+$ for the event $n$. The quantity $p$ corresponds to the purity of the signal.
The function $\varepsilon(m_1^2, m_2^2)$ is the efficiency across the Dalitz plot, and the function $B(m_1^2,m_2^2)$ represents the background in the Dalitz plot. The integrals are computed using Monte Carlo events: since we use varying resonance parameters, the integration is performed at each minimization step.
We minimize the total negative log likelihood
\begin{equation}
\label{eq:likelihoodSum}
\mathcal{F} = \sum_n - 2 \times \log (\mathcal{L}_n),
\end{equation}
where the index $n$ represents a particular event and the sum is performed over the total number of events. We compare the different fits using $\Delta \mathcal{F} = \mathcal{F} - \mathcal{F}_\mathrm{nominal}$, where $\mathcal{F}$ is the value of the total negative log likelihood for a given fit and $\mathcal{F}_\mathrm{nominal}$ is this value for the nominal fit defined below.
We are sensitive only to relative moduli and phases, which means we are free to fix one modulus and one phase. We choose the \DsJRes resonance as the reference amplitude with assignments $|c_{\DsJRes}| = 1$ and $\phi_{\DsJRes} = 0$.

A Monte Carlo simulation is performed using the fit results and is superimposed on the data in the Dalitz plot or on invariant mass projections.
In addition to the value of $\mathcal{F}$, we compute the quantity $\chi^2/\ndof$, where $\ndof$ is the number of degrees of freedom, computed as the number of bins in the Dalitz plot minus the number of free parameters in the fit. We use an adaptive size for the bins so that each bin contains at least 15 events to get an approximately Gaussian behavior.

The fit fraction for each amplitude is defined as:
\begin{equation}
\label{eq:fitFraction}
f_i = \frac{|c_i|^2 \int |A_i|^2 dm_1^2 dm_2^2}{\sum_{j,k} c_j c_k^* \int A_j A_k^* dm_1^2 dm_2^2}.
\end{equation}
The fit fractions do not necessarily add up to 1, due to the interferences that can take place between the different final states. The errors on the fit fractions are evaluated by propagating the full covariance matrix obtained from the fit.

The initial values of the parameters are randomized inside their bounds and 250 different fits are performed with these randomized initial values. We choose the fit which presents the smallest value of the total negative log likelihood $\mathcal{F}$, which allows us to avoid local minima and to obtain the global minimum of the negative log likelihood.

\subsection{Efficiency and background determination}

\noindent
As seen in Eq.~(\ref{eq:likelihood}), the efficiency variation over the Dalitz plot needs to be parametrized. We employ Monte Carlo simulations of the signal events for each reconstructed $D$ decay for the modes \modei and \modexi. The signal density is simulated as being constant over the Dalitz plot. We divide the Dalitz plot in bins of size $0.36\times0.53 {\mathrm{\,(Ge\kern -0.1em V^2\!/}c^4)^2}$. For each two-dimensional bin, we divide the number of simulated signal events after selection and reconstruction by the generated number of events. We combine neighboring bins with low statistics so that each bin has more than ten simulated events after the reconstruction. We obtain a mapping of the efficiency for each reconstructed $D$ mode, and combine these mappings by weighting them together according to the $D$ secondary branching fractions.

To obtain the function $\varepsilon(m_1^2, m_2^2)$ of Eq.~(\ref{eq:likelihood}), we use a bilinear two-dimensional interpolation method applied on the global efficiency mapping. The interpolation makes use of the four values from the bins around a given mass coordinate to compute the efficiency. At the edge of the mapping, we use the value of the bin without interpolation to avoid bias toward a null efficiency.

The background distribution in the Dalitz plot is described by the function $B(m_1^2,m_2^2)$ of Eq.~(\ref{eq:likelihood}). We observe that the Monte Carlo simulation reproduces correctly the data by comparing events in the \mes sideband ($\mes < 5.26 \gevcc$) between the data and the simulation. We therefore employ a Monte Carlo simulation of background events, using the same reconstruction and selection as in the data. We obtain a distribution of the background in the Dalitz plot using bins of size $0.27\times0.37 {\mathrm{\,(Ge\kern -0.1em V^2\!/}c^4)^2}$. The distribution of the background includes the contribution from the combinatorial background (including the background which peaks in the signal region), and we check that the  small proportion of cross-feed events has a similar distribution as the background in the Dalitz plot.
The background distribution is interpolated with a bilinear two-dimensional method to get the value at any coordinate.

\subsection{Fits of the Dalitz plots}

\begin{figure}[htb]
\begin{center}
\epsfig{file=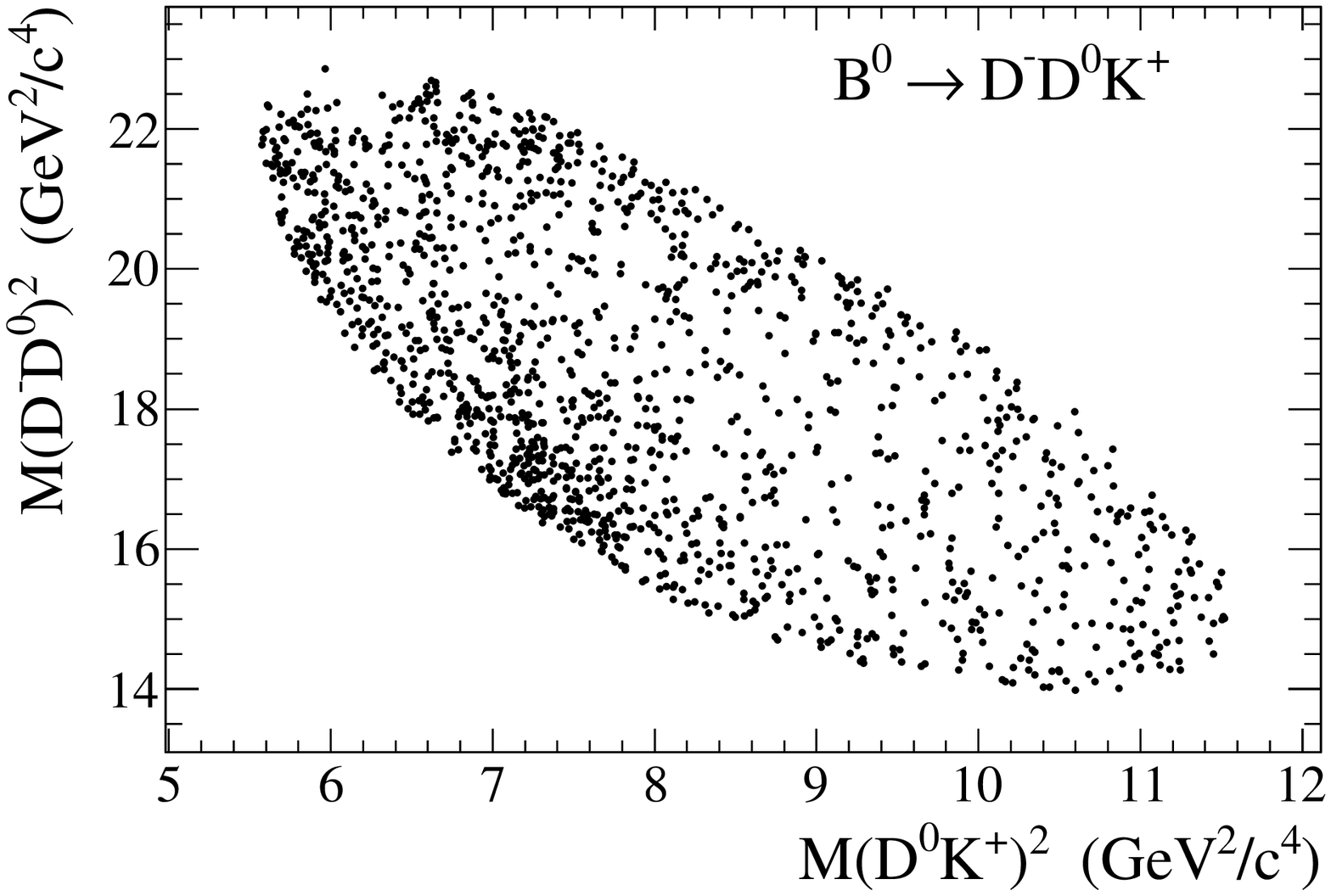,width=7.5cm}
\epsfig{file=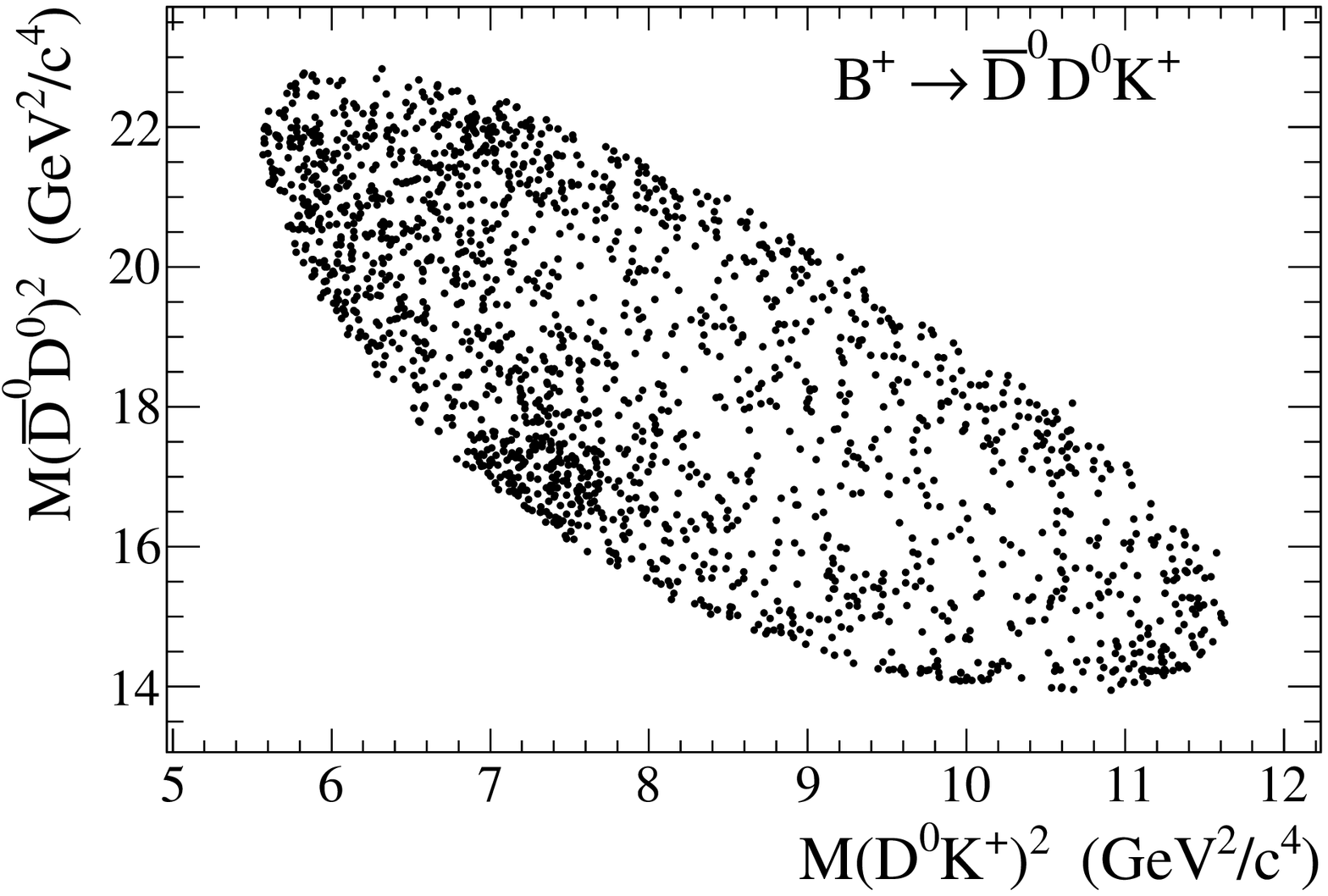,width=7.5cm}
 \caption{Distributions of the data for the Dalitz plot for \modei (top) and \modexi (bottom), where each dot represents a $B$ candidate.}
  \label{fig:dalitz}
\end{center}
\end{figure}

\noindent
The Dalitz plots for \modei and \modexi are shown in Fig.~\ref{fig:dalitz}.
The known amplitudes that could give a contribution in the Dalitz plot for both modes are
\begin{itemize}
  \item nonresonant events,
  \item the \DsJRes meson, and
  \item the \DsTwo meson, which can decay to $D^0K^+$, but has not been observed in $B \to \Dbar^{(*)} D K$ decays.
\end{itemize}
The \DsJTwoEight state decays also to $D^0K^+$ but is not included in the nominal fit.
Furthermore, for the mode \modexi, additional contributions from charmonium states are possible, and are included in the fit:
\begin{itemize}
  \item the \PsiRes meson, and
  \item the $\psi(4160)$ meson.
\end{itemize}
The $\chi_{c2}(2P)$, $\psi(4040)$, and $\psi(4415)$ mesons can also decay to $\Dzb \Dz$. However, they are not included in the nominal fit and are treated separately.
In the following fits, the masses and widths of the resonances are fixed to their world averages~\cite{ref:pdg}, except for the \DsJRes where the parameters are free to vary. The spin of this resonance is assumed to be 1.

\subsubsection{Preliminary fits}

\begin{figure*}[htb]
\begin{center}
\epsfig{file=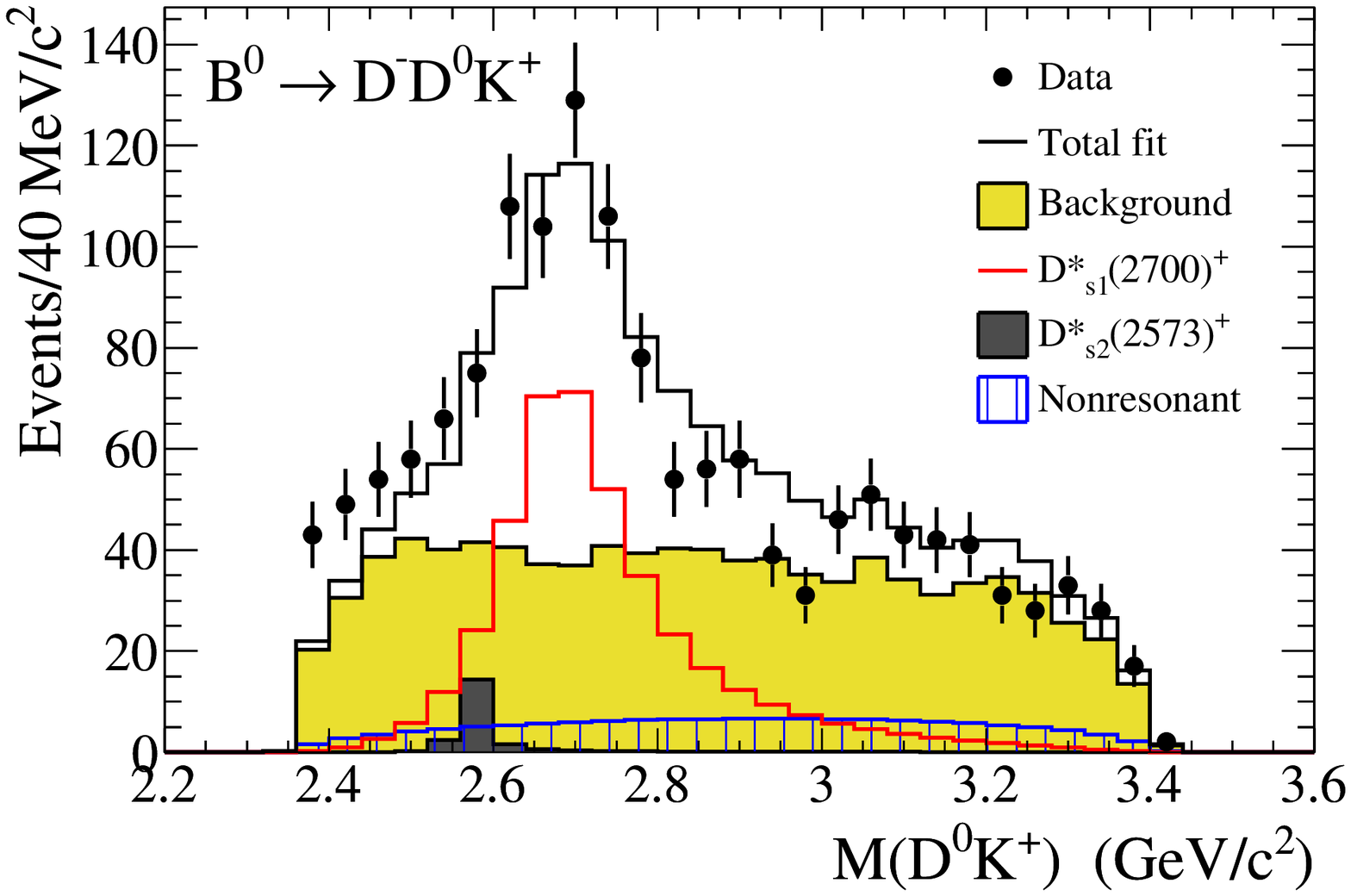,width=8.5cm}
\hspace{0.5cm}
\epsfig{file=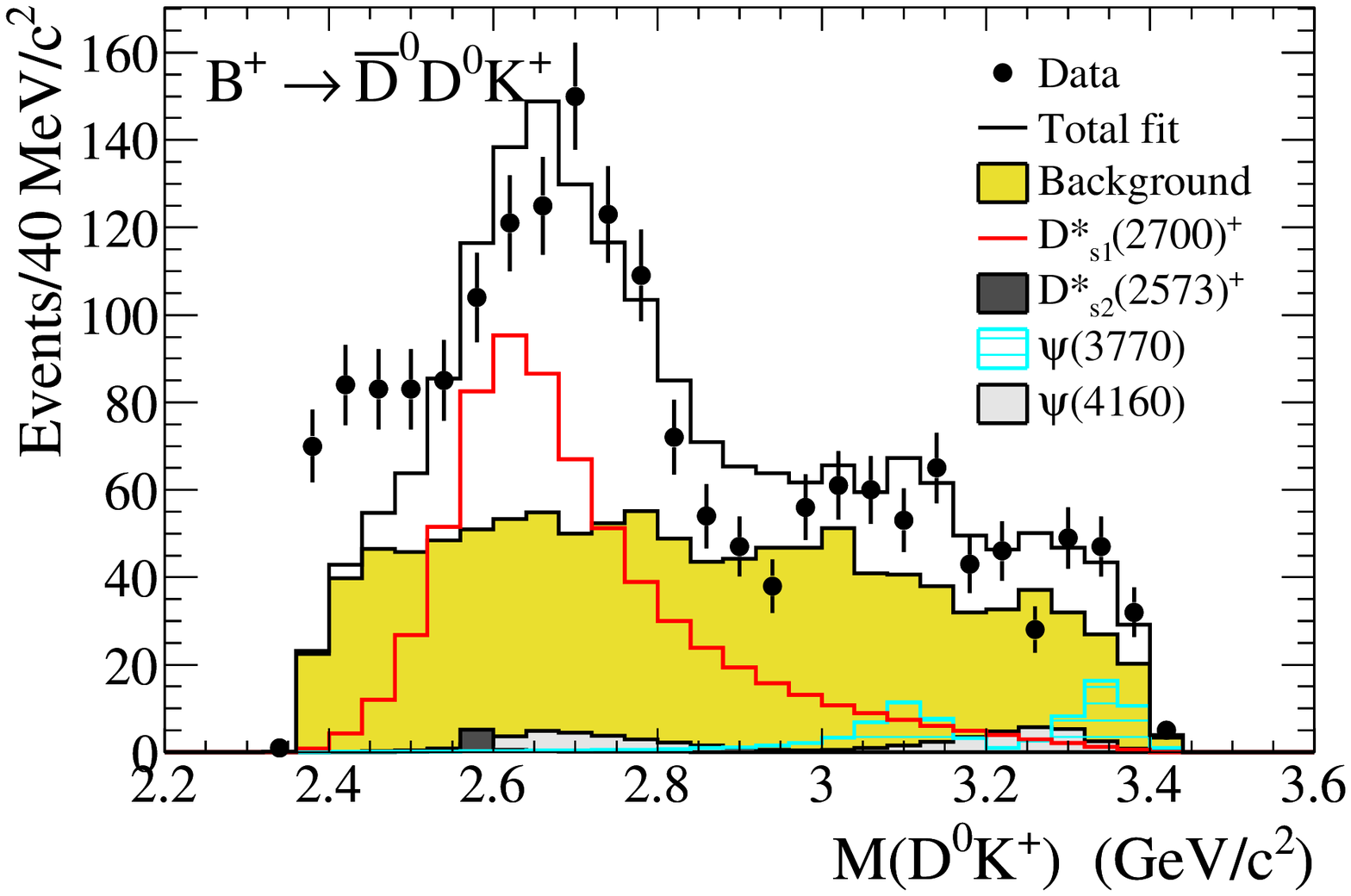,width=8.5cm}
 \caption{Projections of the Dalitz plot on the $D^0K^+$ axis for the data (dots) and for the result of the preliminary fit (total histogram) for the modes \modei (left) and \modexi (right). The fit includes the background (plain yellow histogram), the nonresonant (vertical-hatched blue histogram), the \DsJRes (red histogram), and the \DsTwo (plain dark gray histogram) amplitudes. For \modexi, the additional charmonium contributions consist of the \PsiRes (horizontal-hatched light blue histogram) and the $\psi(4160)$ (plain light gray histogram) amplitudes.}
 \label{fig:dal_nolow}
\end{center}
\end{figure*}

\noindent
We start by a fit to the Dalitz plot with the previously listed amplitudes in the decays \modei and \modexi. The projections on the $D^0K^+$ invariant mass are shown in Fig.~\ref{fig:dal_nolow} (no nonresonant component is included for \modexi as explained below). We see clearly in both cases that the fits are not satisfactory, with values of $\chi^2/\ndof$ (of $\Delta \mathcal{F}$) equal to $82/48$ ($36$) and $265/51$ ($223$) for \modei and \modexi, respectively. In particular, we see that the $D^0K^+$ region between 2350 and 2500 \mevcc is not well described, especially for \modexi. This region corresponds to the threshold in the $D^0K^+$ phase space. The Belle experiment has also reported this enhancement of data with respect to the background in the study of the \modexi decay mode~\cite{ref:DsJBelle}.

\subsubsection{$D^0K^+$ low-mass excess}

\noindent
We check that the enhancement is due to the signal and not the background using two methods to subtract the background. The first method consists of dividing the $D^0K^+$ invariant mass range into $20 \mevcc$ slices and fitting the \mes distribution in each slice. The signal contribution is extracted from the fit, and plotted as a function of mass to obtain a background subtracted $D^0K^+$ invariant mass distribution.
The second method is based on the use of the \emph{sPlot} technique~\cite{ref:splot}, which allows the subtraction of the background in the invariant mass distribution using other uncorrelated variables so that the signal is kept with the correct statistical significance for the variable to be plotted.
The \emph{sPlot} technique uses the results of the \mes fit described in Sec.~\ref{sec:selection}
(yields and covariance matrix) and the PDFs of this fit to compute an event-by-event weight for the signal and background categories. We obtain a $D^0K^+$ invariant mass distribution free of background by applying the \emph{sPlot} weight to each event.
Using these two methods of background subtraction, we observe that the enhancement at low mass is present in the $D^0K^+$ invariant mass distributions for both modes, proving that it originates from $B \to \Db D^0 K^+$ final states.

We verify that this effect cannot be explained by the reflection of a known resonance. A reflection could originate from cross-feed events.
If one of the cross-feed modes of the mode of interest contains a resonance, then this resonance can appear as a structure in the mode of interest, with a smaller magnitude and a shifted mass. As noted earlier, we observe that the cross feed is small for the modes \modei and \modexi. We use simulations of $B \to \Dbar \DsOneRes$, followed by $\DsOneRes \to D^* K$, using the branching fractions from Ref.~\cite{ref:DDKRes} to estimate the possible pollution from these cross-feed modes. We observe that this source of pollution is negligible and cannot explain the enhancement. We also make use of simulations of the cross-feed processes $B \to \Dbar \DsJRes$, followed by $\DsJRes \to D^{(*)} K$, but we observe that these modes give almost no contribution to \modei and \modexi. In conclusion, no reflection seems to explain the enhancement seen at low mass in the $D^0K^+$ invariant mass distributions.

This excess at low mass could be explained by an additional resonance, although none is expected in this mass range.
To test this hypothesis, we refit the data adding the PDF of a scalar resonance with mass and width that are free to float. The quality of the fits improves after the addition of the scalar, with values of $\chi^2/\ndof$ (of $\Delta \mathcal{F}$) equal to $58/44$ ($-2$) and $90/49$ ($-1$) for \modei and \modexi, respectively. The fit for \modei returns a mass and width of the scalar of $2412 \pm 16 \mevcc$ and $163 \pm 64 \mev$ and for \modexi of $2453 \pm 20 \mevcc$ and $283 \pm 45 \mev$, respectively (statistical uncertainties only). These two results are not incompatible ($\sim$1.5$\sigma$ difference for both mass and width, where $\sigma$ is the standard deviation), but the assumption of such a wide resonance at this mass would be speculative. We also try fits with a wide virtual resonance below the $D^0K^+$ threshold, but these fits yield widths with uncertainties that are larger than the corresponding central values. Thus it is not possible to conclude that a real scalar resonance contributes to the $B \to \Db D^0 K^+$ decays. The excess at low mass, although evident, lacks enough statistics to draw definitive conclusions about its nature. In consequence, since the excess at low $D^0K^+$ mass in this data is not understood, we use an arbitrary function to describe it.

This function is chosen to be an exponential starting at the $D^0K^+$ mass threshold.
The exponential function takes the form $A_{\mathrm{Expo}} = e^{-\alpha (m^2_2 - m^2_{2 \, \mathrm{thr}})}$ where $A_{\mathrm{Expo}}$ is the amplitude from Eq.~(\ref{eq:Ai}), $\alpha$ is a free parameter, $m_2$ the $D^0K^+$ invariant mass of the event, and $m_{2 \, \mathrm{thr}}$ the mass threshold, corresponding to the sum of the $D^0$ and $K^+$ masses.
Another approach, introduced in Ref.~\cite{ref:expo}, consists of integrating an exponential contribution as part of the nonresonant amplitude, assuming that the nonresonant amplitude is not necessarily constant over the Dalitz plot.

\subsubsection{Nominal fits}

\noindent
For \modei, the nominal content of the fit includes the nonresonant, the \DsJRes, the \DsTwo, and the exponential amplitudes, which makes a total of nine free parameters in the fit.
For \modexi the fits are not improved when adding the nonresonant component, so we use for the nominal content of the fit the \DsJRes, the \DsTwo, the \PsiRes, the $\psi(4160)$, and the exponential amplitudes. This fit has a total of 11 free parameters.

As stated above, the final fits for \modei and \modexi are each chosen from among 250 fits with randomized initial values of the fit parameters. For \modei, we observe that the majority of the fits (60\%) converge to the exact same values of the set of fitted parameters, which we choose as our nominal fit. However a proportion of fits (23\%) converges to a slightly lower value of $\mathcal{F}$ with a sum of fit fractions (Eq.~(\ref{eq:fitFraction})) greater than 250\%. This large sum of fractions, which is unlikely to be physical, originates from large interference between the nonresonant and the exponential contributions, which are both \emph{ad~hoc} amplitudes. These fits return values of the parameter related to the resonances very close to those of the nominal fit: these differences are taken into account in the calculation of the systematic uncertainties. Moreover, we find a local minimum for 3\% of the fits with $\Delta \mathcal{F} = 1$, which we account for in the systematic uncertainties. We observe that 5\% of the fits have $\Delta \mathcal{F} > 20$. For \modexi, one-third of the 250 fits converge to the global minimum, chosen as our nominal fit. We observe four local minima, representing a proportion of 42\% of the fits, situated at $\Delta \mathcal{F} < 4$ of the global minimum. Since these fits are close to the nominal one, we use them as a contribution to the systematic uncertainties for the parameters related to the resonances. We see that 10\% of the fits have $\Delta \mathcal{F} > 20$, justifying the use of the procedure of randomizing the initial values of the parameters.

\begin{figure*}[htb]
\begin{center}
\epsfig{file=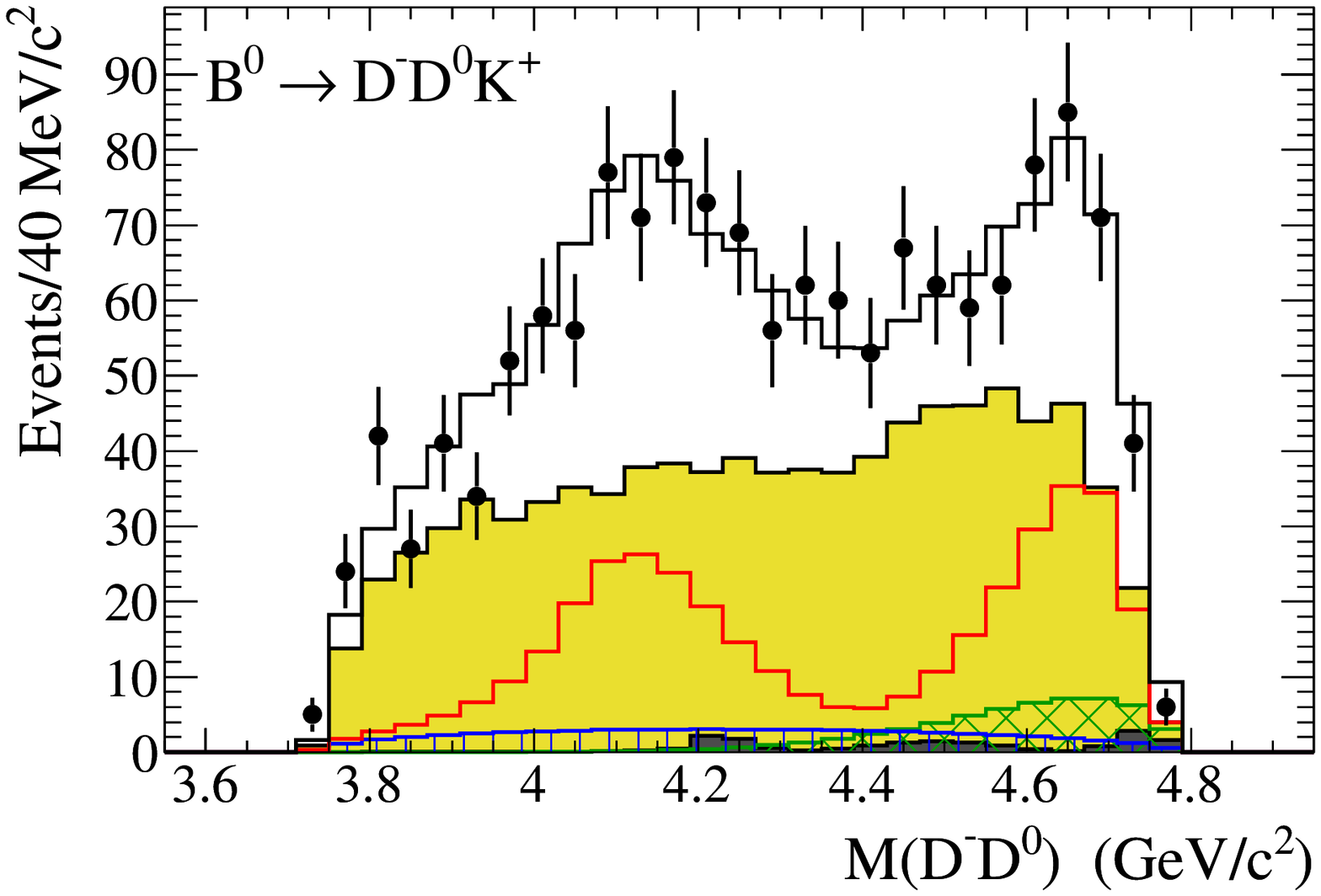,width=8.5cm}
\hspace{0.5cm}
\epsfig{file=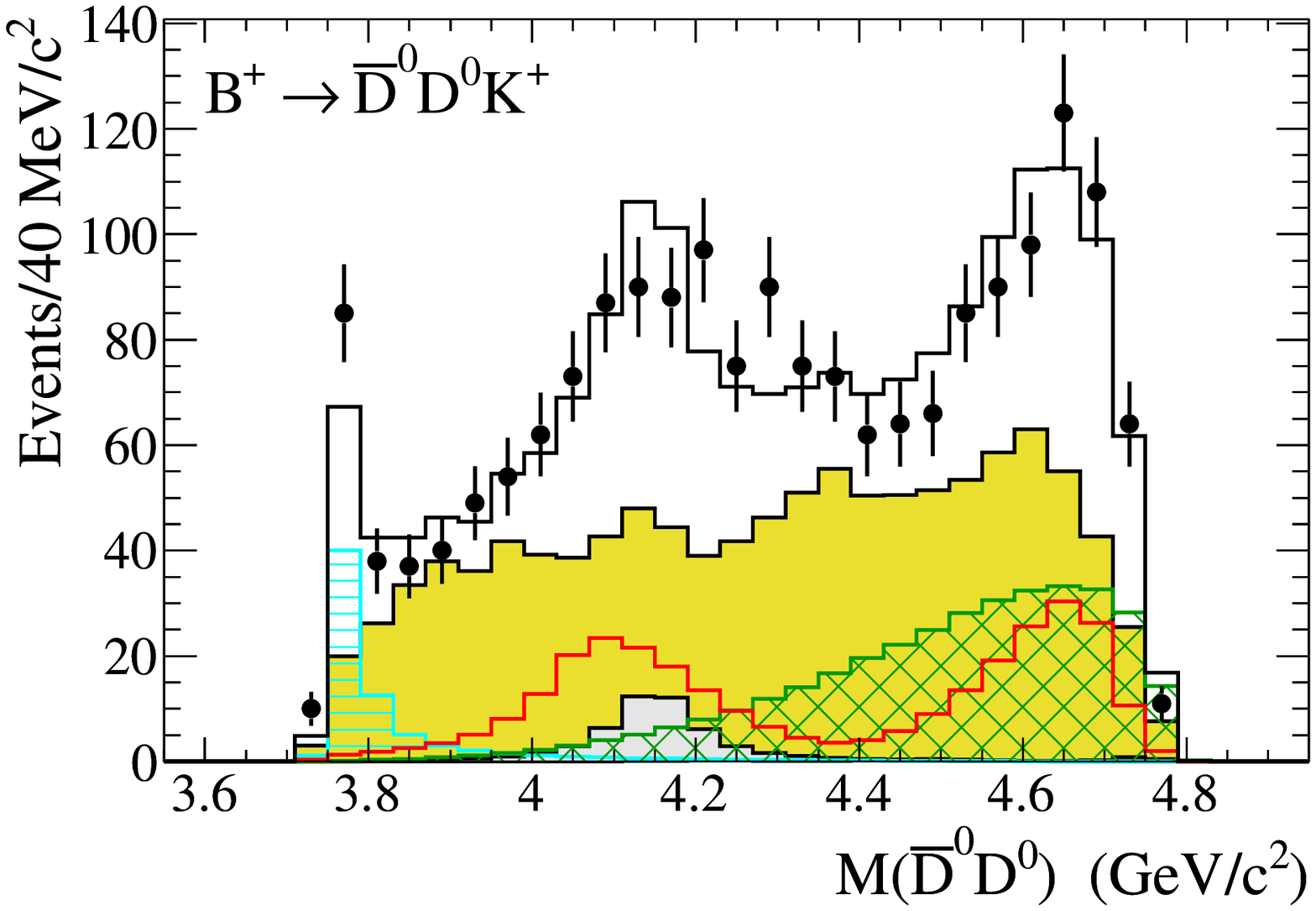,width=8.5cm}
\epsfig{file=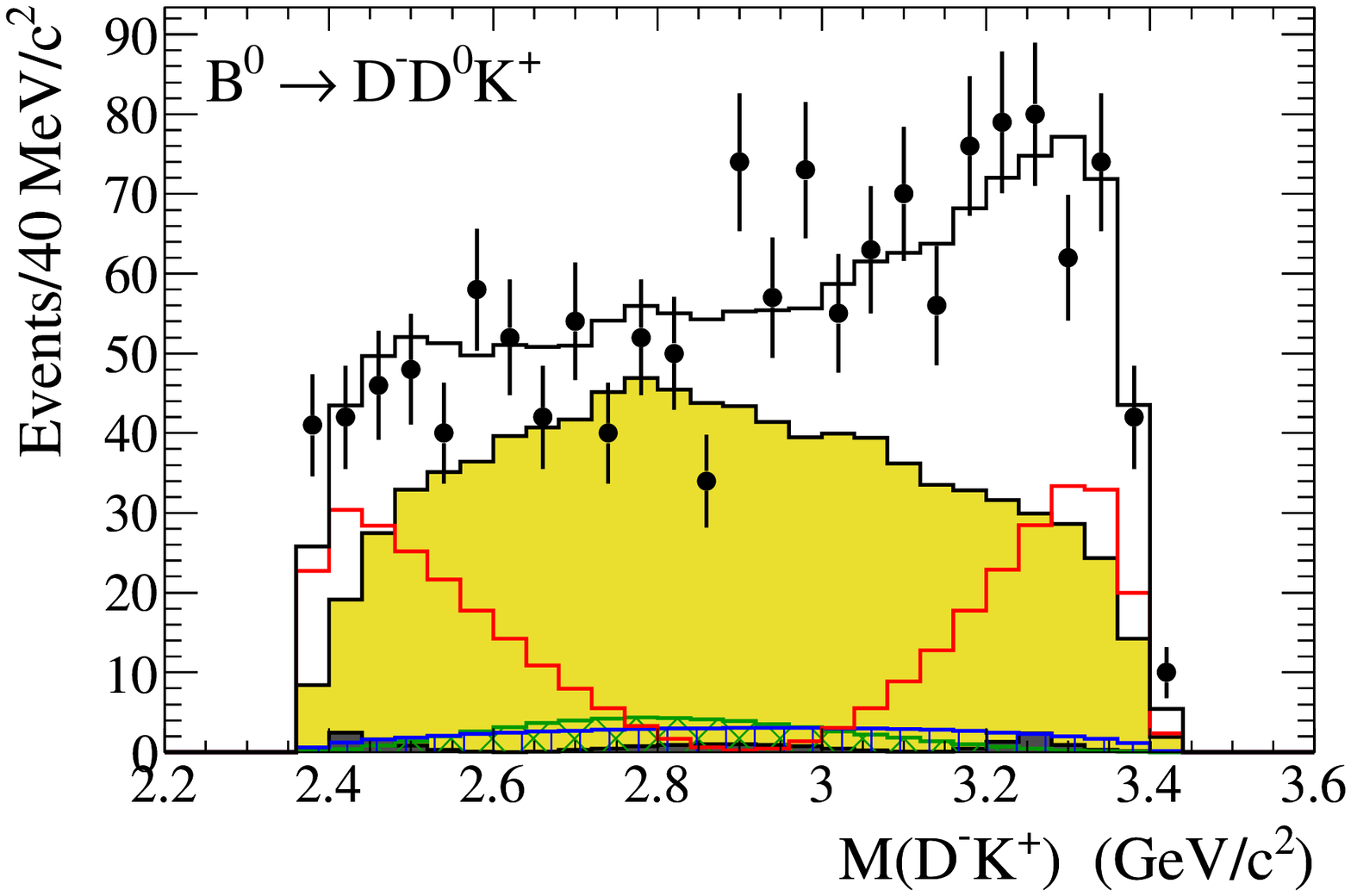,width=8.5cm}
\hspace{0.5cm}
\epsfig{file=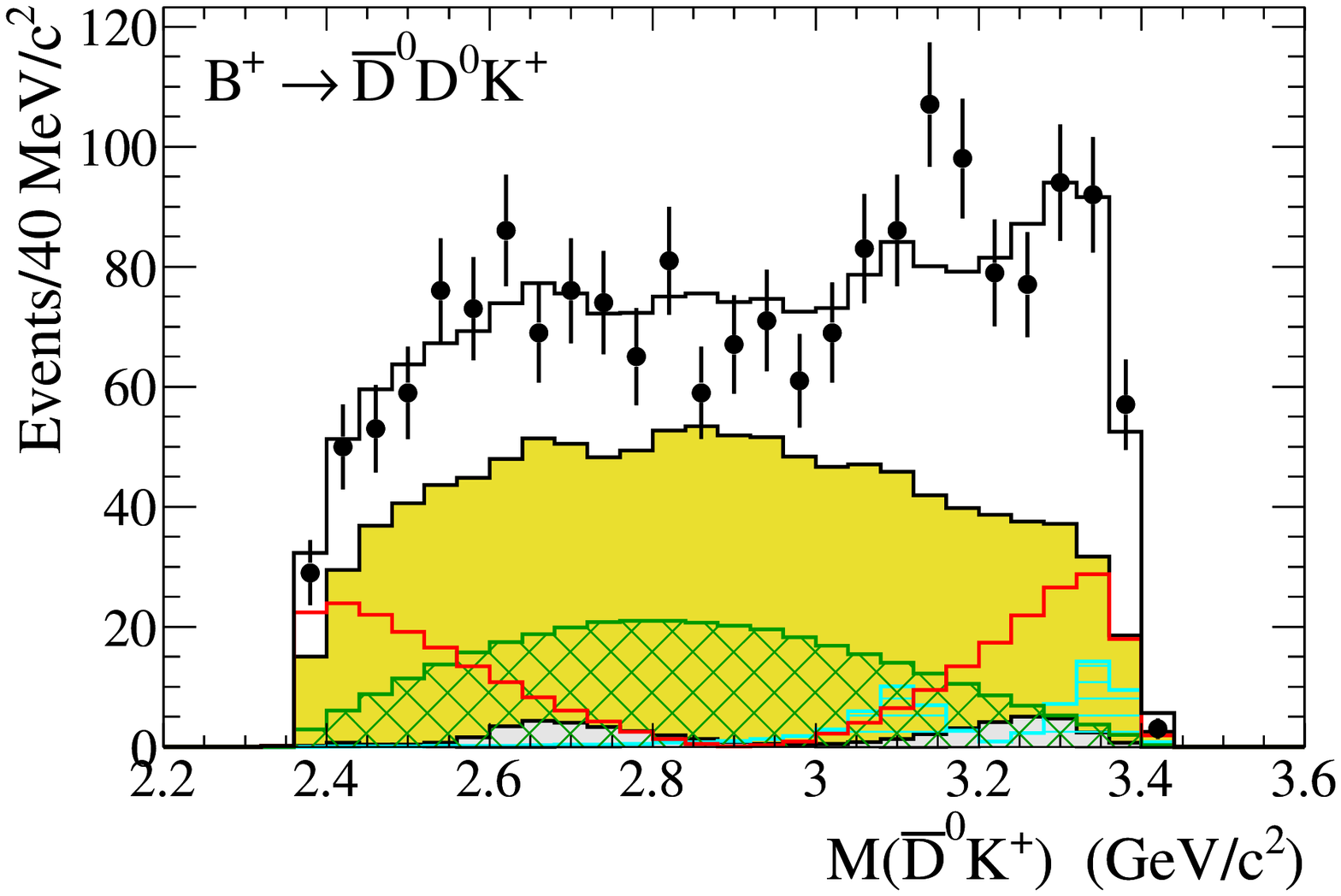,width=8.5cm}
\epsfig{file=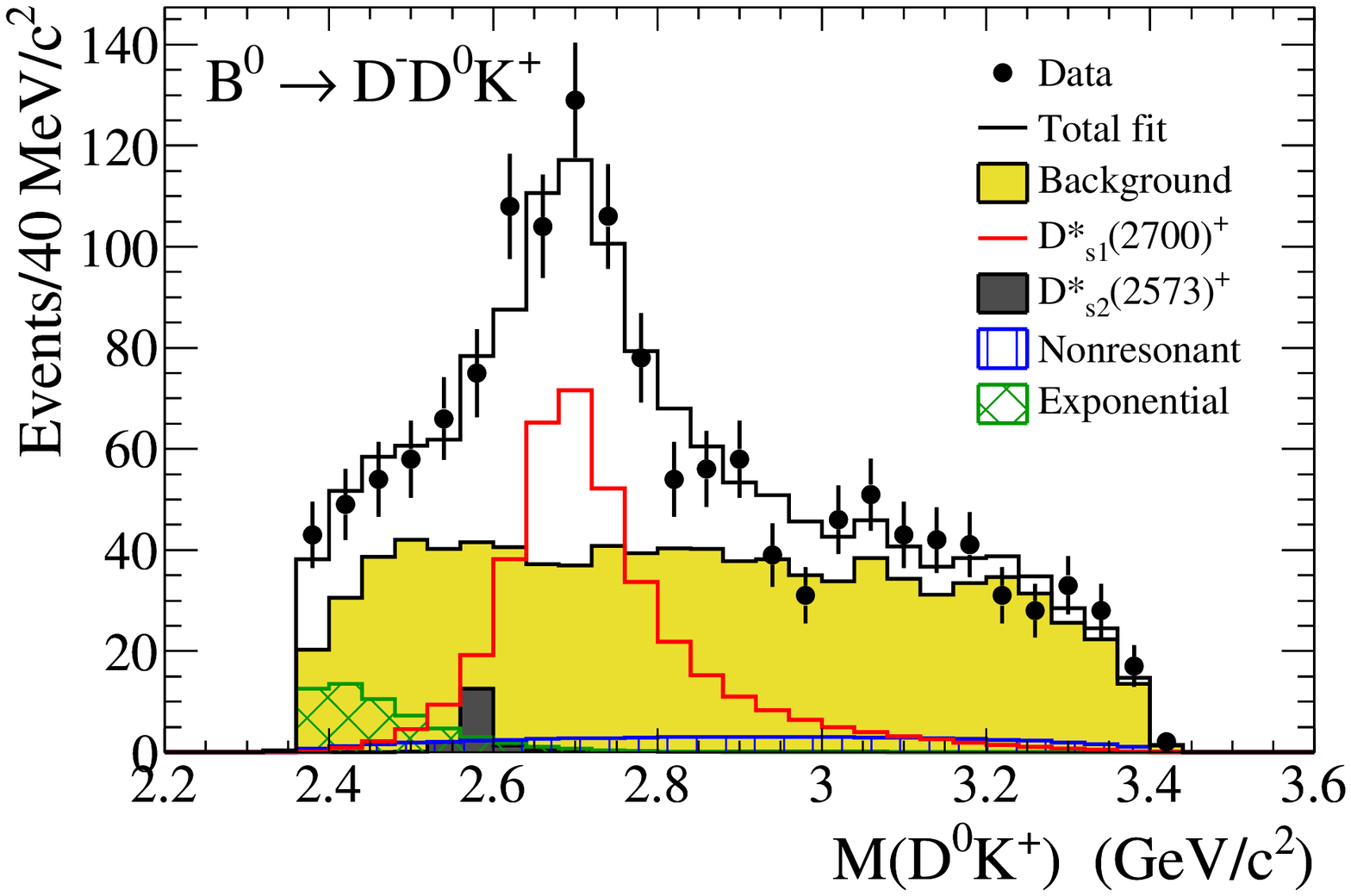,width=8.5cm}
\hspace{0.5cm}
\epsfig{file=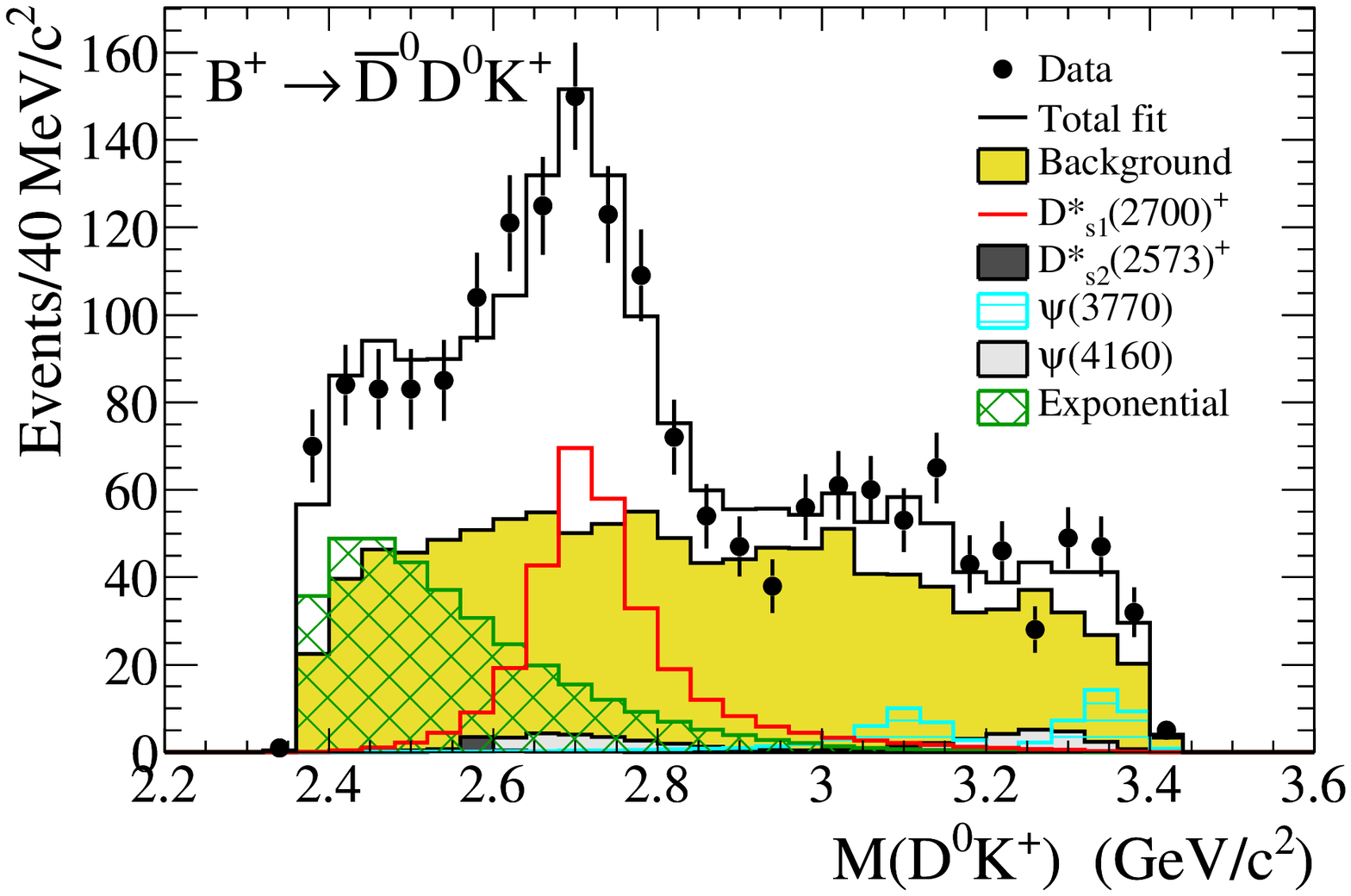,width=8.5cm}
 \caption{Projections of the Dalitz plot on the three axes for the data (dots) and for the result of the nominal fit (total histogram) for the modes \modei (left) and \modexi (right). The fit includes the background (plain yellow histogram), the nonresonant (vertical-hatched blue histogram, present only for \modei), the \DsJRes (red histogram), the \DsTwo (plain dark gray histogram), and the exponential (cross-hatched green histogram) amplitudes. For \modexi, the additional charmonium contributions consist of the \PsiRes (horizontal-hatched light blue histogram) and the $\psi(4160)$ (plain light gray histogram) amplitudes.}
 \label{fig:dal}
\end{center}
\end{figure*}

The nominal fit for \modei is shown in Fig.~\ref{fig:dal} and returns $\chi^2/\ndof=56/45$. The nominal fit for \modexi is presented in Fig.~\ref{fig:dal} and gives $\chi^2/\ndof=86/48$. The high value of the $\chi^2/\ndof$ can be partly explained by some discrepancies between the data and our fit located mainly in the \PsiRes region. We do not extract any information from this region, and we consider that the fit gives a satisfactory description of the data in other regions.

To assess the values found for $\mathcal{F}$ for the nominal fit for each mode, we generate a large number of Monte Carlo samples based on the PDFs of the nominal fits with the statistics of the data and fit these samples with the same method as in the data. We observe that the nominal values of $\mathcal{F}$ for the two modes are close ($0.2\sigma$) to the central values of the distributions of $\mathcal{F}$ obtained from the Monte Carlo samples. Similarly to the data, the simulations show the presence of several local minima close to the global minimum.

Comparing the fit results before and after removing a resonance with fixed shape parameters allows us to translate directly the difference of negative log likelihood as a $\chi^2$ distribution with two degrees of freedom (modulus and phase). A difference $\Delta \mathcal{F}$ of 12 and 29 corresponds roughly to a statistical significance of $3\sigma$ and $5\sigma$, respectively. This significance does not take into account the systematic uncertainties, and the final significance decreases after taking them into account. This property is used here to estimate the need for the \DsTwo and $\psi(4160)$ resonances.
In a first stage, we repeat the nominal fit without the \DsTwo amplitude. We observe that the minimum log likelihood increases with $\Delta \mathcal{F} = 16$ for \modei and $\Delta \mathcal{F} = 5$ for \modexi, indicating that the presence of the \DsTwo resonance is significant in \modei. Removing the $\psi(4160)$ component from the Dalitz plot fit of \modexi gives $\Delta \mathcal{F} = 23$.

Adding an additional amplitude for either the \DsJTwoEight or the \DsJThreeZero resonance does not improve the fits (assuming a spin of 1 for these two states).
For \modexi, adding a contribution for either the $\chi_{c2}(2P)$, $\psi(4040)$, or $\psi(4415)$ meson does not improve the fit to a significant level. None of the parameters changed by a statistically significant amount when adding to the fit these extra resonances, and no systematic error is assigned for these resonances.

\section{Systematic uncertainties}

\begin{table*}[htb]
 \begin{center}
 \caption{Systematic uncertainties for \modei in the moduli, in the phases ($^\circ$), in the fractions (\%), in the mass (MeV/$c^2$) and the width (MeV) of the \DsJRes, and in $\alpha$, the parameter of the exponential function. The labels refer to systematic uncertainties related to the fit bias (``Bias"), efficiency interpolation (``Eff."), statistical uncertainty in efficiency (``Eff. II"), background knowledge (``Bkg"), Blatt-Weisskopf barrier (``BW"), low-mass description (``Low mass"), \DsTwo amplitude (``$D^*_{s2}$"), and other minima (``Min.''). The sign `---' means that no systematic uncertainty has been attributed to the specific parameter.}
 \label{tab:syst1}
  \vskip 0.2cm
 \begin{tabular}{lrrrrrrrrrr}
 \hline
 \hline
 \CellTopTwo
 Parameter & ~~Value~~ & ~~Bias~~ & ~~Eff.~~ & ~~Eff. II~~ & ~~Bkg~~ & ~~BW~~ & ~~Low~ & ~~$D^*_{s2}$~~ & ~~Min.~~ & ~~Total~~ \\
 & & &  & & & & mass & & &  \\
 \hline
  \CellTopThree
Modulus \DsTwo & 0.031 & $\pm 0.001$ & $0$ & $\pm 0.001$ & $0$ & $\pm 0.001$ & $0$ & $0$ & $\pm 0.001$ & $\pm 0.002$ \\
 \CellTopThree
Modulus nonresonant & 1.33 & $\pm 0.09$ & $\pm 0.03$ & $\pm 0.05$ & ${}^{+0.21}_{-0.33}$ & ${}^{+0.39}_{-0.00}$ & $0$ & $+0.04$ & --- & ${}^{+0.46}_{-0.35}$ \\
 \CellTopThree
Modulus exponential & 6.94 & $\pm 0.24$ & $\pm 0.06$ & $\pm 0.19$ & ${}^{+0.12}_{-0.21}$ & ${}^{+0.72}_{-0.20}$ & $0$ & $+0.21$ & --- & ${}^{+0.82}_{-0.43}$ \\
 \CellTopThree
Phase \DsTwo& 277 & $\pm 4$ & $\pm 1$ & $\pm 2$ & $\pm 2$ & ${}^{+0}_{-8}$ & ${}^{+2}_{-0}$ & $0$ & ${}^{+4}_{-0}$ & ${}^{+6}_{-9}$ \\
 \CellTopThree
Phase nonresonant& 287 & $\pm 2$ & $0$ & $\pm 1$ & ${}^{+9}_{-6}$ & ${}^{+2}_{-14}$ & $0$ & $0$ & --- & ${}^{+10}_{-15}$ \\
 \CellTopThree
Phase exponential& 269 & $\pm 6$ & $\pm 1$ & $\pm 4$ & ${}^{+3}_{-0}$ & ${}^{+0}_{-13}$ & $0$ & $+15$ & --- & ${}^{+17}_{-15}$ \\
\CellTopThree
Fraction \DsJRes& 66.7 & --- & $\pm 0.2$ & $\pm 0.6$ & ${}^{+2.2}_{-2.1}$ & ${}^{+0.4}_{-2.1}$ & ${}^{+1.3}_{-0.0}$ & $+0.6$ & ${}^{+2.3}_{-2.3}$ & ${}^{+3.5}_{-3.8}$ \\
 \CellTopThree
Fraction \DsTwo& 3.2 & --- & $0$ & $\pm 0.2$ & ${}^{+0.0}_{-0.1}$ & ${}^{+0.0}_{-0.3}$ & ${}^{+0.2}_{-0.0}$ & $0$ & ${}^{+0.1}_{-0.0}$ & ${}^{+0.3}_{-0.4}$ \\
 \CellTopThree
Fraction nonresonant& 10.9 & --- & $\pm 0.3$ & $\pm 0.7$ & ${}^{+3.3}_{-4.3}$ & ${}^{+6.1}_{-0.0}$ & $0$ & $+0.2$ & --- & ${}^{+7.0}_{-4.3}$ \\
 \CellTopThree
Fraction exponential& 9.9 & --- & $\pm 0.2$ & $\pm 0.5$ & ${}^{+2.9}_{-1.5}$ & ${}^{+0.0}_{-2.9}$ & $0$ & $+0.9$ & --- & ${}^{+3.0}_{-3.3}$ \\
 \CellTopThree
$M(\DsJRes)$ & 2694 & $\pm 2$ & $0$ & $\pm 1$ & $0$ & ${}^{+13}_{-2}$ & ${}^{+0}_{-1}$ & $0$ & ${}^{+3}_{-0}$ & ${}^{+13}_{-3}$ \\
 \CellTopThree
$\Gamma(\DsJRes)$ & 145 & $\pm 8$ & $\pm 1$ & $\pm 3$ & ${}^{+4}_{-3}$ & ${}^{+17}_{-9}$ & ${}^{+5}_{-0}$ & $-6$ & ${}^{+11}_{-4}$ & ${}^{+22}_{-14}$ \\
 \CellTopThree
$\alpha$ & 1.43 & $\pm 0.11$ & $\pm 0.02$ & $\pm 0.08$ & ${}^{+0.20}_{-0.26}$ & ${}^{+0.48}_{-0.00}$ & $0$ & $-0.06$ & --- & ${}^{+0.54}_{-0.30}$ \\
\hline
\hline
 \end{tabular}
 \end{center}
\end{table*}

\begin{table*}[htb]
 \begin{center}
 \caption{Systematic uncertainties for \modexi in the moduli, in the phases ($^\circ$), in the fractions (\%), in the mass (MeV/$c^2$) and the width (MeV) of the \DsJRes, and in $\alpha$, the parameter of the exponential function. The labels refer to systematic uncertainties related to the fit bias (``Bias"), efficiency interpolation (``Eff."), statistical uncertainty in efficiency (``Eff. II"), background knowledge (``Bkg"), Blatt-Weisskopf barrier (``BW"), low-mass description (``Low mass"), \DsTwo amplitude (``$D^*_{s2}$"), \PsiRes amplitude (``$\psi$"), and other minima (``Min.''). The sign `---' means that no systematic uncertainty has been attributed to the specific parameter.}
 \label{tab:syst11}
  \vskip 0.2cm
 \begin{tabular}{lrrrrrrrrrrr}
 \hline
 \hline
 \CellTopTwo
 Parameter & ~~Value~~ & ~~Bias~~ & ~~Eff.~~ & ~~Eff. II~~ & ~~Bkg~~ & ~~BW~~ & ~~Low~ & ~~$D^*_{s2}$~~ & ~~$\psi$~~ & ~~Min.~~ & ~~Total~~ \\
 & & &  & & & & mass & & & & \\
 \hline
  \CellTopThree
Modulus \DsTwo & 0.021 & $\pm 0.001$ & $\pm 0.001$ & $\pm 0.001$ & $0$ & ${}^{+0.003}_{-0.000}$ & ${}^{+0.006}_{-0.000}$ & $0$ & $0$ & ${}^{+0.005}_{-0.002}$ & ${}^{+0.009}_{-0.003}$ \\
 \CellTopThree
Modulus \PsiRes & 1.40 & $\pm 0.08$ & $\pm 0.02$ & $\pm 0.03$ & ${}^{+0.05}_{-0.06}$ & ${}^{+0.03}_{-0.10}$ & ${}^{+0.06}_{-0.00}$ & $+0.07$ & $-0.19$ & ${}^{+0.14}_{-0.00}$ & ${}^{+0.20}_{-0.24}$ \\
 \CellTopThree
Modulus $\psi(4160)$ & 0.78 & $\pm 0.02$ & $\pm 0.01$ & $\pm 0.03$ & $\pm 0.01$ & ${}^{+0.00}_{-0.10}$ & ${}^{+0.16}_{-0.00}$ & $+0.06$ & $-0.02$ & ${}^{+0.06}_{-0.08}$ & ${}^{+0.18}_{-0.14}$ \\
 \CellTopThree
Modulus exponential & 16.15 & $\pm 0.53$ & $\pm 0.13$ & $\pm 0.35$ & ${}^{+0.57}_{-0.70}$ & ${}^{+0.00}_{-1.44}$ & $0$ & $+0.66$ & $-0.18$ & --- & ${}^{+1.09}_{-1.74}$ \\
 \CellTopThree
Phase \DsTwo& 267 & $\pm 9$ & $\pm 3$ & $\pm 3$ & ${}^{+7}_{-8}$ & ${}^{+1}_{-3}$ & ${}^{+5}_{-0}$ & $0$ & $0$ & ${}^{+10}_{-0}$ & ${}^{+17}_{-13}$ \\
 \CellTopThree
Phase \PsiRes& 284 & $\pm 5$ & $0$ & $\pm 2$ & $\pm 1$ & ${}^{+0}_{-22}$ & ${}^{+7}_{-19}$ & $-2$ & $-4$ & ${}^{+25}_{-4}$ & ${}^{+26}_{-30}$ \\
 \CellTopThree
Phase $\psi(4160)$& 188 & $\pm 6$ & $0$ & $\pm 2$ & $\pm 4$ & ${}^{+0}_{-3}$ & ${}^{+12}_{-0}$ & $-1$ & $+1$ & ${}^{+2}_{-15}$ & ${}^{+14}_{-17}$ \\
 \CellTopThree
Phase exponential& 308 & $\pm 2$ & $\pm 1$ & $\pm 2$ & $\pm 2$ & $\pm 4$ & $0$ & $+2$ & $+1$ & --- & ${}^{+6}_{-5}$ \\
\CellTopThree
Fraction \DsJRes& 38.3 & --- & $\pm 0.3$ & $\pm 0.6$ & ${}^{+0.3}_{-0.0}$ & ${}^{+0.0}_{-0.1}$ & ${}^{+0.0}_{-2.4}$ & $-1.4$ & $0$ & ${}^{+0.0}_{-5.5}$ & ${}^{+0.8}_{-6.2}$ \\
 \CellTopThree
Fraction \DsTwo& 0.6 & --- & $\pm 0.1$ & $\pm 0.1$ & $0$ & ${}^{+0.2}_{-0.0}$ & ${}^{+0.3}_{-0.0}$ & $0$ & $0$ & ${}^{+0.2}_{-0.1}$ & ${}^{+0.4}_{-0.2}$ \\
 \CellTopThree
Fraction \PsiRes& 9.0 & --- & $0$ & $\pm 0.3$ & ${}^{+0.3}_{-0.2}$ & ${}^{+0.0}_{-0.5}$ & ${}^{+0.1}_{-0.4}$ & $+0.1$ & $-0.4$ & $\pm 0.1$ & ${}^{+0.4}_{-0.8}$ \\
 \CellTopThree
Fraction $\psi(4160)$& 6.4 & --- & $\pm 0.1$ & $\pm 0.3$ & $\pm 0.3$ & ${}^{+0.0}_{-0.7}$ & ${}^{+1.7}_{-0.1}$ & $+0.4$ & $-0.1$ & ${}^{+0.7}_{-2.3}$ & ${}^{+1.9}_{-2.4}$ \\
 \CellTopThree
Fraction exponential& 44.5 & --- & $\pm 0.2$ & $\pm 0.6$ & ${}^{+0.0}_{-0.2}$ & ${}^{+0.0}_{-2.0}$ & $0$ & $+1.1$ & $-0.2$ & --- & ${}^{+1.3}_{-2.1}$ \\
 \CellTopThree
$M(\DsJRes)$ & 2707 & $\pm 4$ & $0$ & $\pm 1$ & $\pm 3$ & ${}^{+7}_{-2}$ & ${}^{+0}_{-4}$ & $+1$ & $0$ & ${}^{+0}_{-5}$ & $\pm 8$ \\
 \CellTopThree
$\Gamma(\DsJRes)$ & 113 & $\pm 5$ & $\pm 1$ & $\pm 3$ & ${}^{+9}_{-7}$ & ${}^{+17}_{-0}$ & ${}^{+0}_{-9}$ & $-5$ & $+2$ & ${}^{+0}_{-7}$ & ${}^{+20}_{-16}$ \\
 \CellTopThree
$\alpha$ & 0.68 & $\pm 0.01$ & $0$ & $\pm 0.01$ & $\pm 0.01$ & ${}^{+0.02}_{-0.00}$ & $0$ & $-0.02$ & $0$ & --- & $\pm 0.02$ \\
\hline
\hline
 \end{tabular}
 \end{center}
\end{table*}

\noindent
We consider several sources of systematic uncertainties in the fit parameters, such as the moduli, the phases, the fit fractions, and the mass and width of the \DsJRes.
Tables~\ref{tab:syst1} and~\ref{tab:syst11} give the details of the systematic uncertainties.

To estimate a potential fit bias, we generate a large number of Monte Carlo samples based on the nominal fits with the same statistics as in the data. We perform the Dalitz plot fit on these samples and extract the central value and uncertainty of each parameter. We obtain a distribution of the pull for each parameter, defined as the difference between the central value from a particular Monte Carlo sample and the nominal value, divided by the uncertainty in the fit value from the Monte Carlo sample. These distributions have a width compatible with~1 as expected, but their mean is observed to be shifted with respect to~0, which points toward fit biases that we correct for in our results. The systematic uncertainty related to this bias correction is taken as half of the bias added in quadrature
with the uncertainty in the bias (the systematic uncertainty related to the bias is labeled as ``Bias" in Tables~\ref{tab:syst1} and~\ref{tab:syst11}).

Systematic uncertainties arise from the efficiency calculation. To estimate these we first use the raw efficiency in each bin of the Dalitz plot instead of the interpolation, and consider the difference for each parameter as an estimate of the systematic uncertainties for the efficiency (labeled as ``Eff." in Tables~\ref{tab:syst1} and~\ref{tab:syst11}). Second, we take into account the statistical fluctuation on the efficiency due to the finite number of Monte Carlo events. In each bin of the Dalitz plot, we randomize the efficiency within its statistical uncertainty and produce many new mappings of the efficiency. The analysis is performed using these new efficiency mappings, which gives a distribution for each parameter from which we extract the systematic uncertainties (column ``Eff. II").

Another source of systematic uncertainty comes from the background description. Repeating the analysis using the raw value of the background in each bin of the Dalitz plot instead of the interpolation does not change the results. In addition, the signal purity is varied according to its total uncertainty (see Sec.~\ref{sec:selection}), which allows us to get the systematic uncertainty related to the signal and background knowledge (column ``Bkg").

Several systematic uncertainties arise from the fit modeling. The first one comes from the Blatt-Weisskopf barrier, which is not known precisely.
Fits of both modes with this value as a free parameter show that the analysis is not sensitive to it.
To estimate the related systematic uncertainty,
we repeat the analysis changing the value of the Blatt-Weisskopf barrier radius from the nominal value 1.5 GeV$^{-1}$ to 5 GeV$^{-1}$ and 0 GeV$^{-1}$. The differences for each parameter between the nominal fit and these fits give the systematic uncertainties (labeled as ``BW" in the tables). Another systematic uncertainty originates from the description of the low-mass excess. Since this effect is not understood in the current data, it is important to estimate the possible influence it induces on the fit parameters, especially the mass and the width of the \DsJRes meson. To compute the systematic uncertainties associated with this effect, we assume first that the excess originates from a scalar resonance at low mass: we repeat the fits replacing the exponential contribution by a scalar resonance with a mass and a width free to float. Second, instead of the exponential function, we use an alternative model, $A_{\mathrm{Alt}} = \frac{1}{1 + a \times (m^2_2 - m^2_{2 \, \mathrm{thr}})}$ with $a$ a free parameter. The maximum deviations with respect to the nominal fit for each of these two alternatives are used as the systematic uncertainties (column ``Low mass").
We then study the influence of the resonances on the analyses.
Since the \DsTwo amplitude presents a low fit fraction, we repeat the fits without this amplitude and take the difference as a systematic uncertainty (column ``$D^*_{s2}$").
The effect of the spin-1 \DsJTwoEight has also been investigated: because the fits return negligible fractions for this state, the systematic uncertainties on the other parameters are found to be negligible.
For \modexi, another systematic uncertainty (column ``$\psi$") arises from the \PsiRes\ parameters fixed to the world average value~\cite{ref:pdg}. As an alternative, we use the measurement from the KEDR experiment~\cite{ref:KEDR}. This experiment reports a result which takes into account the resonance-continuum interference in the near-threshold region and which agrees well with previous \babar measurements~\cite{ref:DDKRes,psiBaBar}.

As discussed previously, 250 fits are performed for each mode with randomized initial values of the fit parameters, which allows us to find the nominal fit. We find several minima which are close to the nominal fit, and we use the largest differences in parameter values (for those related to resonances) as contributions to their systematic uncertainties (column ``Min.").

\section{Results}

\subsection{Dalitz plot analysis}

\noindent
The results for the Dalitz plot analysis of the modes \modei and \modexi are presented in Tables~\ref{tab:result1} and~\ref{tab:result11}.
In both modes, we observe a large contribution of the \DsJRes resonance via $\Bz\to \Dm \DsJRes, \DsJRes \to \Dz \Kp$ for \modei and via $\Bu \to \Dzb \DsJRes, \DsJRes \to \Dz \Kp$ for \modexi.
This is the first time the \DsJRes is observed in the decay \modei. We observe that the \DsTwo meson contributes a small fraction to \modei whereas it is not significant in the other mode. We note that the \DsTwo meson is expected to be suppressed in \BToDDK decays due to its spin value of 2. We observe the decay $\Bu \to \PsiRes \Kp, \PsiRes \to \Dzb \Dz$, which confirms one of our previous results~\cite{ref:DDKRes}. We notice that the process $\Bu \to \psi(4160) \Kp$ followed by $\psi(4160) \to \Dzb \Dz$ is needed to improve the description of the data. The low-mass excess in the $D^0K^+$ invariant mass is evident in \modei and is the main contribution in \modexi. With this data sample it is not possible to determine the origin of this excess. The exponential function used to describe the effect has a parameter $\alpha$ equal to $1.43 \pm 0.71 {}^{+0.54}_{-0.30}$ and $0.68 \pm 0.08 \pm 0.02$ for \modei and \modexi, respectively.

\begin{table*}[htb]
 \begin{center}
 \caption{Results from the Dalitz plot fit (moduli, phases, and fractions) for \modei. The different contributions are listed: the \DsJRes and \DsTwo resonances, the nonresonant amplitude, and the low-mass excess described by an exponential. The first uncertainties are statistical and the second systematic.}
 \label{tab:result1} \vskip 0.2cm
 \begin{tabular}{lrrr}
 \hline
 \hline
 \CellTop
Contribution & Modulus & Phase ($^\circ$) & ~Fraction (\%) \\
\hline
\CellTopTwo
\DsJRes & 1.00 & 0 & $66.7 \pm 7.8 {}^{+3.5}_{-3.8}$ \\
\CellTopTwo
\DsTwo & $0.031 \pm 0.008 \pm 0.002$ & $277 \pm 17 {}^{+6}_{-9}$ & $3.2 \pm 1.6 {}^{+0.3}_{-0.4}$ \\
\CellTopTwo
Nonresonant & $1.33 \pm 0.63 {}^{+0.46}_{-0.35}$ & $287 \pm 21 {}^{+10}_{-15}$ & $10.9 \pm 6.6 {}^{+7.0}_{-4.3}$ \\
\CellTopTwo
Exponential & $6.94 \pm 1.83 {}^{+0.82}_{-0.43}$ & $269 \pm 33 {}^{+17}_{-15}$ & $9.9 \pm 2.9 {}^{+3.0}_{-3.3}$ \\
\hline
\CellTopTwo
Sum & & & $90.6 \pm 10.7 {}^{+8.4}_{-6.7}$ \\
\hline
\hline
 \end{tabular}
 \end{center}
\end{table*}

\begin{table*}[htb]
 \begin{center}
 \caption{Results from the Dalitz plot fit (moduli, phases, and fractions) for \modexi. The different contributions are listed: the \DsJRes, \DsTwo, \PsiRes, and $\psi(4160)$ resonances, and the low-mass excess described by an exponential. The first uncertainties are statistical and the second systematic.}
 \label{tab:result11} \vskip 0.2cm
 \begin{tabular}{lrrr}
 \hline
 \hline
 \CellTopTwo
Contribution & Modulus & Phase ($^\circ$) & ~Fraction (\%) \\
 \hline
 \CellTopTwo
\DsJRes & 1.00 & 0 & $38.3 \pm 5.0 {}^{+0.8}_{-6.2}$ \\
 \CellTopTwo
\DsTwo & $0.021 \pm 0.010 {}^{+0.009}_{-0.003}$ & $267 \pm 30 {}^{+17}_{-13}$ & $0.6 \pm 1.1 {}^{+0.4}_{-0.2}$ \\
 \CellTopTwo
\PsiRes & $1.40 \pm 0.21 {}^{+0.20}_{-0.24}$ & $284 \pm 22 {}^{+26}_{-30}$ & $9.0 \pm 3.1 {}^{+0.4}_{-0.8}$ \\
 \CellTopTwo
$\psi(4160)$ & $0.78 \pm 0.20 {}^{+0.18}_{-0.14}$ & $188 \pm 13 {}^{+14}_{-17}$ & $6.4 \pm 3.1 {}^{+1.9}_{-2.4}$ \\
 \CellTopTwo
Exponential & $16.15 \pm 2.26 {}^{+1.09}_{-1.74}$ & $308 \pm 8 {}^{+6}_{-5}$ & $44.5 \pm 6.2 {}^{+1.3}_{-2.1}$ \\
\hline
 \CellTopTwo
Sum & & & $98.9 \pm 9.2 {}^{+2.5}_{-7.0}$ \\
\hline
\hline
 \end{tabular}
 \end{center}
\end{table*}

\subsection{Branching fractions}

\noindent
The partial branching fraction $\BR_\mathrm{res}$ for a given resonance is computed using the fraction $f_\mathrm{res}$ of the resonance (see Tables~\ref{tab:result1} and~\ref{tab:result11}) and the total branching fraction $\BR_\mathrm{tot}$ of the specific $B$ mode. We use the total branching fractions measured in a previous publication~\cite{ref:DDKBF} with the exact same data sample. The computation is as follows:
\begin{equation}
\BR_\mathrm{res} = f_\mathrm{res} \times \BR_\mathrm{tot}. \nonumber
\end{equation}

The uncertainties for the partial branching fraction are computed from the quadratic sum of the uncertainties from the total branching fraction~\cite{ref:DDKBF} and the uncertainties from the fraction (see Tables~\ref{tab:result1} and \ref{tab:result11}), where we treat separately the statistical and systematic uncertainties.

The results are presented in Table~\ref{tab:summaryBF}. We can compare these results with previous publications that are available for the mode \modexi. In \babar~\cite{ref:DDKRes}, using an analysis of the $\Dzb \Dz$ invariant mass, the result for the \PsiRes\ was $\BR(B^{+} \to \PsiRes K^{+} \; [\Dbar^{0}D^{0}]) = (1.41 \pm 0.30 \pm 0.22) \times 10^{-4}$, which is in good agreement with the current result. This present measurement supersedes the previous one. The Belle experiment~\cite{ref:DsJBelle} finds for the partial branching fraction $\BR(B^{+} \to \Dbar^{0} \DsJRes \; [D^{0}K^{+}]) = (11.3 \pm 2.2 {}_{-2.8}^{+1.4}) \times 10^{-4}$ and $\BR(B^{+} \to \PsiRes K^{+} \; [\Dbar^{0}D^{0}]) = (2.2 \pm 0.5 \pm 0.3) \times 10^{-4}$, which present a difference with our measurements of $1.7\sigma$ and $1.4\sigma$, respectively.

The significance of the decay of the $\psi(4160)$ charmonium resonance to $\Dbar^{0}D^{0}$ is $3.3\sigma$, including systematic uncertainties. The significance of the \DsTwo meson decay to $D^{0}K^{+}$ is $3.4\sigma$ (for the mode \modei), including systematic uncertainties.

\begin{table}[htb]
 \begin{center}
 \caption{Summary of partial branching fractions. The first uncertainties are statistical and the second systematic. The notation $B^{0} \to D^{-} \DsJRes \; [D^{0}K^{+}]$ refers to $B^{0} \to D^{-} \DsJRes$ followed by $\DsJRes \to D^{0}K^{+}$.}
 \label{tab:summaryBF}
  \vskip 0.2cm
 \begin{tabular}{ll}
 \hline
 \hline
 \CellTop
 Mode & \BR \, $(10^{-4})$ \\
 \hline
 \CellTop
$B^{0} \to D^{-} \DsJRes \; [D^{0}K^{+}]$ ~~~~~~~& $7.14 \pm 0.96 \pm 0.69$\\
$B^{+} \to \Dbar^{0} \DsJRes \; [D^{0}K^{+}]$ & $5.02 \pm 0.71 \pm 0.93$\\
$B^{0} \to D^{-} \DsTwo \; [D^{0}K^{+}]$ & $0.34 \pm 0.17 \pm 0.05$\\
$B^{+} \to \Dbar^{0} \DsTwo \; [D^{0}K^{+}]$ & $0.08 \pm 0.14 \pm 0.05$\\
$B^{+} \to \PsiRes K^{+} \; [\Dbar^{0}D^{0}]$ & $1.18 \pm 0.41 \pm 0.15$\\
$B^{+} \to \psi(4160) K^{+} \; [\Dbar^{0}D^{0}]$ & $0.84 \pm 0.41 \pm 0.33$\\
\hline
\hline
 \end{tabular}
 \end{center}
\end{table}

\subsection{Properties of \DsJRes}

\begin{table}[htb]
 \begin{center}
 \caption{Mass and width of the \DsJRes\ meson obtained from the Dalitz plot analyses of the modes \modei and \modexi. The first uncertainties are statistical and the second systematic.}
 \label{tab:allDsJ}
  \vskip 0.2cm
 \begin{tabular}{lll}
 \hline
 \hline
 \CellTop
Mode~~~~~~~~~~~~~~~~~ & Mass (MeV/$c^2$)~~~~ & Width (MeV) \\
 \hline
 \CellTop
 \modei & $2694 \pm 8 {}^{+13}_{-3}$ & $145 \pm 24 {}^{+22}_{-14}$ \\
\CellTopTwo
\modexi & $2707 \pm 8 \pm 8$ & $113 \pm 21 {}^{+20}_{-16}$ \\
\hline
\hline
 \end{tabular}
 \end{center}
\end{table}

\noindent
We show in Table~\ref{tab:allDsJ} the result for the mass and width of the \DsJRes\ meson for the two modes obtained from the Dalitz plot analysis. The measurements in the two final states agree with each other within their uncertainties.
We combine the measurements for the mass and width, respectively, calculating the weighted means and taking into account the asymmetric uncertainties. 
This procedure works for uncertainties that are not correlated between the measurements. Only the systematic uncertainty coming from the Blatt-Weisskopf factor is correlated between the modes: we perform first the combination without this particular systematic uncertainty, where we obtain $M(\DsJRes) = 2699 \pm 7 \mevcc$ and $\Gamma(\DsJRes) = 127 \pm 17 \mev$ (including statistical and systematic). For the uncertainty related to the Blatt-Weisskopf factor, we use the maximum positive and negative deviations found in the two modes that we add quadratically to the total uncertainties.

Finally, the combination of both modes gives :
\begin{eqnarray}
M(\DsJRes) &=& 2699 {}^{+14}_{-7} \mevcc,  \\
\Gamma(\DsJRes) &=& 127 {}^{+24}_{-19} \mev, \nonumber
\end{eqnarray}
where the uncertainties quoted are the total uncertainties (including statistical and systematic).

These values can be compared to the current world average of $M(\DsJRes) = 2709 \pm 4 \mevcc$ and
$\Gamma(\DsJRes) = 117 \pm 13 \mev$~\cite{ref:pdg}. Our measurements are compatible with the world averages.

The Dalitz plot analyses have been performed with the spin hypothesis $J=1$ for the \DsJRes. To test this hypothesis, we repeat the fits using the hypotheses $J=0$ and $J=2$. The results are presented in Table~\ref{tab:spin}. We conclude from this table that $J=0,2$ are not able to fit correctly the data, and that $J=1$ is strongly favored. Because of parity conservation, we deduce that the \DsJRes\ meson is a state with $J^P=1^-$, which confirms the measurement performed by the Belle experiment~\cite{ref:DsJBelle}.

\begin{table*}[htb]
 \begin{center}
 \caption{Value of $\Delta \mathcal{F}$ and $\chi^2/\ndof$ for the hypotheses $J=0,1,2$ for the two modes. The nominal fit is presented in bold characters.}
 \label{tab:spin}
  \vskip 0.2cm
 \begin{tabular}{lllclll}
 \hline
 \hline
 \CellTop
Mode & \multicolumn{2}{c}{$J=0$} & \multicolumn{1}{c}{$J=1$} & \multicolumn{2}{c}{$J=2$} \\
& $\Delta \mathcal{F}$~~ & $\chi^2/\ndof$ & ~~~~~~$\chi^2/\ndof$~~~~~~ & $\Delta \mathcal{F}$~~ & $\chi^2/\ndof$ \\
 \hline
 \CellTop
\modei ~~~~~~~~ & $131$ & $131/45$ & \textbf{56/45} & $108$ & 125/45 \\
\CellTop
\modexi & $63$ & 137/48 & \textbf{86/48} & $99$ & 145/48 \\
\hline
\hline
 \end{tabular}
 \end{center}
\end{table*}

\section{Conclusions}

\noindent
We have analyzed $471 \times 10^6$ pairs of $B$ mesons recorded by the \babar experiment and studied the decays \modei and \modexi. Dalitz plot analyses of these decays have been performed, where we extract moduli and phases for each contribution to these decay modes. We observe the \DsJRes meson in both final states and measure its mass and width to be
\begin{eqnarray}
M(\DsJRes) &=& 2699 {}^{+14}_{-7} \mevcc,  \\
\Gamma(\DsJRes) &=& 127 {}^{+24}_{-19} \mev, \nonumber
\end{eqnarray}
where the uncertainties include statistical and systematic uncertainties. We determine its spin parity to be $J^P=1^-$.

Several possibilities have been discussed to interpret the \DsJRes meson.
This resonance could be interpreted as the $n^{2S+1}L_J = 1^3D_1$ $c \bar{s}$ state~\cite{ref:DsJTheoOverlap} or as the first radial excitation of the $D_s^*(2112)$ meson ($2^3S_1$)~\cite{ref:DsJTheoRadial}.
Some authors interpret the \DsJRes state as a mixing between the $2^3S_1$ and the $1^3D_1$ states~\cite{ref:DsJTheoMixing}, obtaining a model which is able to solve some of the discrepancies with the experimental data.
Another possibility would be that the signal interpreted as the \DsJRes originates from two resonances overlapping each other~\cite{ref:DsJTheoOverlap}.

We observe an enhancement between 2350 and 2500 \mevcc in the $D^0K^+$ invariant mass that we are not able to interpret. This enhancement, which has an important contribution in the \modexi decay mode, is described by an \emph{ad~hoc} function in the Dalitz plot fit. This effect was also seen in the Belle experiment in the study of the \modexi final state~\cite{ref:DsJBelle}.

It is not clear what could be the cause of this enhancement in the $D^0K^+$ invariant mass, although we note that some $D_s$ excited states are expected in this mass range in some models~\cite{ref:DsJTheoRadial,ref:DsJPred} and have not been observed yet.
Some authors, as for example in Ref.~\cite{ref:expo}, who have observed the same effect in other channels claim that it could be a specific form of a nonresonant amplitude.

Finally, we do not observe the \DsJTwoEight and \DsJThreeZero resonances in the final states \modei and \modexi.

\section*{Acknowledgments}
\noindent
We are grateful for the
extraordinary contributions of our \pep2\ colleagues in
achieving the excellent luminosity and machine conditions
that have made this work possible.
The success of this project also relies critically on the
expertise and dedication of the computing organizations that
support \babar.
The collaborating institutions wish to thank
SLAC for its support and the kind hospitality extended to them.
This work is supported by the
US Department of Energy
and National Science Foundation, the
Natural Sciences and Engineering Research Council (Canada),
the Commissariat \`a l'Energie Atomique and
Institut National de Physique Nucl\'eaire et de Physique des Particules
(France), the
Bundesministerium f\"ur Bildung und Forschung and
Deutsche Forschungsgemeinschaft
(Germany), the
Istituto Nazionale di Fisica Nucleare (Italy),
the Foundation for Fundamental Research on Matter (The Netherlands),
the Research Council of Norway, the
Ministry of Education and Science of the Russian Federation,
Ministerio de Econom\'{\i}a y Competitividad (Spain), the
Science and Technology Facilities Council (United Kingdom),
and the Binational Science Foundation (U.S.-Israel).
Individuals have received support from
the Marie-Curie IEF program (European Union) and the A. P. Sloan Foundation (USA).




\begin{thebibliography}{99}
\bibitem{cc}
{Charge conjugate reactions are assumed throughout this article.}
\end{thebibliography}
\end{document}